\documentclass[final]{IEEEtran}
\usepackage{amsthm,amssymb,graphicx,multirow,amsmath,color,amsfonts}
\usepackage[update,prepend]{epstopdf}
\usepackage[noadjust]{cite}
\usepackage[latin1]{inputenc}
\usepackage{tikz}
\usepackage{bm}
\usepackage{bbm} 
\usepackage[nolist]{acronym}
\usepackage{pdfpages}
\usepackage{multirow}
\usepackage{subfig}
\usepackage{comment}
\usepackage{fancyhdr}

\def\nb0{{\mathbf{0}}}
\def\nb1{{\mathbf{1}}}

\def\P{\mathbb{P}}

\allowdisplaybreaks 

\usepackage{setspace}	

\begin{acronym}

\acro{5G-NR}{5G New Radio}
\acro{3GPP}{3rd Generation Partnership Project}
\acro{AC}{address coding}
\acro{ACF}{autocorrelation function}
\acro{ACR}{autocorrelation receiver}
\acro{ADC}{analog-to-digital converter}
\acrodef{aic}[AIC]{Analog-to-Information Converter}     
\acro{AIC}[AIC]{Akaike information criterion}
\acro{aric}[ARIC]{asymmetric restricted isometry constant}
\acro{arip}[ARIP]{asymmetric restricted isometry property}

\acro{ARQ}{automatic repeat request}
\acro{AUB}{asymptotic union bound}
\acrodef{awgn}[AWGN]{Additive White Gaussian Noise}     
\acro{AWGN}{additive white Gaussian noise}

\acro{APSK}[PSK]{asymmetric PSK} 

\acro{waric}[AWRICs]{asymmetric weak restricted isometry constants}
\acro{warip}[AWRIP]{asymmetric weak restricted isometry property}
\acro{BCH}{Bose, Chaudhuri, and Hocquenghem}        
\acro{BCHC}[BCHSC]{BCH based source coding}
\acro{BEP}{bit error probability}
\acro{BFC}{block fading channel}
\acro{BG}[BG]{Bernoulli-Gaussian}
\acro{BGG}{Bernoulli-Generalized Gaussian}
\acro{BPAM}{binary pulse amplitude modulation}
\acro{BPDN}{Basis Pursuit Denoising}
\acro{BPPM}{binary pulse position modulation}
\acro{BPSK}{binary phase shift keying}
\acro{BPZF}{bandpass zonal filter}
\acro{BSC}{binary symmetric channels}              
\acro{BU}[BU]{Bernoulli-uniform}
\acro{BER}{bit error rate}
\acro{BS}{base station}

\acro{CP}{Cyclic Prefix}
\acrodef{cdf}[CDF]{cumulative distribution function}   
\acro{CDF}{cumulative distribution function}
\acrodef{c.d.f.}[CDF]{cumulative distribution function}
\acro{CCDF}{complementary cumulative distribution function}
\acrodef{ccdf}[CCDF]{complementary CDF}               
\acrodef{c.c.d.f.}[CCDF]{complementary cumulative distribution function}
\acro{CD}{cooperative diversity}

\acro{CDMA}{Code Division Multiple Access}
\acro{ch.f.}{characteristic function}
\acro{CIR}{channel impulse response}
\acro{cosamp}[CoSaMP]{compressive sampling matching pursuit}
\acro{CR}{cognitive radio}
\acro{cs}[CS]{compressed sensing}                   
\acrodef{cscapital}[CS]{Compressed sensing}
\acrodef{CS}[CS]{compressed sensing}
\acro{CSI}{channel state information}

\acro{CCSDS}{consultative committee for space data systems}
\acro{CC}{convolutional coding}
\acro{CRB}{Cram{\'e}r-Rao bound}
\acro{BCRB}{Bayesian CRB}

\acro{DAA}{detect and avoid}
\acro{DAB}{digital audio broadcasting}
\acro{DCT}{discrete cosine transform}
\acro{dft}[DFT]{discrete Fourier transform}
\acro{DR}{distortion-rate}
\acro{DS}{direct sequence}
\acro{DS-SS}{direct-sequence spread-spectrum}
\acro{DTR}{differential transmitted-reference}
\acro{DVB-H}{digital video broadcasting\,--\,handheld}
\acro{DVB-T}{digital video broadcasting\,--\,terrestrial}
\acro{DL}{downlink}
\acro{DSSS}{Direct Sequence Spread Spectrum}
\acro{DFT-s-OFDM}{Discrete Fourier Transform-spread-Orthogonal Frequency Division Multiplexing}
\acro{DAS}{distributed antenna system}
\acro{DNA}{Deoxyribonucleic Acid}

\acro{EC}{European Commission}
\acro{EED}[EED]{exact eigenvalues distribution}
\acro{EIRP}{Equivalent Isotropically Radiated Power}
\acro{ELP}{equivalent low-pass}
\acro{eMBB}{Enhanced Mobile Broadband}
\acro{EMF}{electric and magnetic fields}
\acro{EU}{European union}

\acro{FC}[FC]{fusion center}
\acro{FCC}{Federal Communications Commission}
\acro{FEC}{forward error correction}
\acro{FFT}{fast Fourier transform}
\acro{FH}{frequency-hopping}
\acro{FH-SS}{frequency-hopping spread-spectrum}
\acrodef{FS}{Frame synchronization}
\acro{FSsmall}[FS]{frame synchronization}  
\acro{FDMA}{Frequency Division Multiple Access} 
\acro{FIM}{Fisher information matrix}

\acro{GA}{Gaussian approximation}
\acro{GF}{Galois field }
\acro{GG}{Generalized-Gaussian}
\acro{GIC}[GIC]{generalized information criterion}
\acro{GLRT}{generalized likelihood ratio test}
\acro{GPS}{Global Positioning System}
\acro{GMSK}{Gaussian minimum shift keying}
\acro{GSMA}{Global System for Mobile communications Association}

\acro{HAP}{high altitude platform}

\acro{IDR}{information distortion-rate}
\acro{IFFT}{inverse fast Fourier transform}
\acro{iht}[IHT]{iterative hard thresholding}
\acro{i.i.d.}{independent, identically distributed}
\acro{IoT}{Internet of Things}                      
\acro{IR}{impulse radio}
\acro{lric}[LRIC]{lower restricted isometry constant}
\acro{lrict}[LRICt]{lower restricted isometry constant threshold}
\acro{ISI}{intersymbol interference}
\acro{ITU}{International Telecommunication Union}
\acro{ICNIRP}{International Commission on Non-Ionizing Radiation Protection}
\acro{IEEE}{Institute of Electrical and Electronics Engineers}
\acro{ICES}{IEEE international committee on electromagnetic safety}
\acro{IEC}{international Electrotechnical Commission}
\acro{IARC}{International Agency on Research on Cancer}
\acro{IS-95}{Interim Standard 95}
\acro{ISAC}{integrated sensing and communications}

\acro{KPI}{Key Performance Indicator}
\acro{KLD}{Kullback-Leibler divergence}

\acro{LEO}{low earth orbit}
\acro{LF}{likelihood function}
\acro{LLF}{log-likelihood function}
\acro{LLR}{log-likelihood ratio}
\acro{LLRT}{log-likelihood ratio test}
\acro{LoS}{line-of-sight}
\acro{LRT}{likelihood ratio test}
\acro{wlric}[LWRIC]{lower weak restricted isometry constant}
\acro{wlrict}[LWRICt]{LWRIC threshold}
\acro{LPWAN}{low power wide area network}
\acro{LoRaWAN}{Low power long Range Wide Area Network}

\acro{MB}{multiband}
\acro{MC}{multicarrier}
\acro{MDS}{mixed distributed source}
\acro{MF}{matched filter}
\acro{m.g.f.}{moment generating function}
\acro{MI}{mutual information}
\acro{MIMO}{multiple-input multiple-output}
\acro{MISO}{multiple-input single-output}
\acrodef{maxs}[MJSO]{maximum joint support cardinality}                       
\acro{ML}[ML]{maximum likelihood}
\acro{MMSE}{minimum mean squared error}
\acro{MSE}{mean squared error}
\acro{MMV}{multiple measurement vectors}
\acrodef{MOS}{model order selection}
\acro{M-PSK}[${M}$-PSK]{$M$-ary phase shift keying}                       
\acro{M-APSK}[${M}$-PSK]{$M$-ary asymmetric PSK} 

\acro{M-QAM}[$M$-QAM]{$M$-ary quadrature amplitude modulation}
\acro{MRC}{maximal ratio combiner}                  
\acro{maxs}[MSO]{maximum sparsity order}                                      
\acro{M2M}{machine to machine}                                                
\acro{MUI}{multi-user interference}
\acro{mMTC}{massive Machine Type Communications}      
\acro{mm-Wave}[mm-Wave]{millimeter-wave}
\acro{MP}{mobile phone}
\acro{MPE}{maximum permissible exposure}
\acro{MAC}{media access control}

\acro{NB}{narrowband}
\acro{NBI}{narrowband interference}
\acro{NLA}{nonlinear sparse approximation}
\acro{NLOS}{Non-Line of Sight}
\acro{NTIA}{National Telecommunications and Information Administration}
\acro{NTP}{National Toxicology Program}

\acro{OC}{optimum combining}                             
\acro{OC}{optimum combining}
\acro{ODE}{operational distortion-energy}
\acro{ODR}{operational distortion-rate}
\acro{OFDM}{orthogonal frequency-division multiplexing}
\acro{omp}[OMP]{orthogonal matching pursuit}
\acro{OSMP}[OSMP]{orthogonal subspace matching pursuit}
\acro{OQAM}{offset quadrature amplitude modulation}
\acro{OQPSK}{offset QPSK}
\acro{OFDMA}{Orthogonal Frequency-division Multiple Access}

\acro{OQPSK/PM}{OQPSK with phase modulation}

\acro{PAM}{pulse amplitude modulation}
\acro{PAR}{peak-to-average ratio}                 
\acro{PDF}{probability density function}
\acrodef{p.d.f.}[PDF]{probability distribution function}
\acro{PDP}{power dispersion profile}
\acro{PMF}{probability mass function}                             
\acrodef{p.m.f.}[PMF]{probability mass function}
\acro{PN}{pseudo-noise}
\acro{PPM}{pulse position modulation}
\acro{PRake}{Partial Rake}
\acro{PSD}{power spectral density}
\acro{PSEP}{pairwise synchronization error probability}
\acro{PSK}{phase shift keying}
\acro{PD}{power density}
\acro{8-PSK}[$8$-PSK]{$8$-phase shift keying}
\acro{PHP}{Poisson hole process}
 \acro{PPP}{Poisson point process}
\acrodefplural{PPP}{Poisson point processes}
\acro{FSK}{frequency shift keying}

\acro{QAM}{Quadrature Amplitude Modulation}
\acro{QPSK}{quadrature phase shift keying}
\acro{OQPSK/PM}{OQPSK with phase modulator }

\acro{RD}[RD]{raw data}
\acro{RDL}{"random data limit"}
\acro{ric}[RIC]{restricted isometry constant}
\acro{rict}[RICt]{restricted isometry constant threshold}
\acro{rip}[RIP]{restricted isometry property}
\acro{ROC}{receiver operating characteristic}
\acro{rq}[RQ]{Raleigh quotient}
\acro{RS}[RS]{Reed-Solomon}
\acro{RSC}[RSSC]{RS based source coding}
\acro{r.v.}{random variable}                               
\acro{R.V.}{random vector}
\acro{RMS}{root mean square}
\acro{RFR}{radiofrequency radiation}
\acro{RIS}{reconfigurable intelligent surface}

\acro{SA}[SA-Music]{subspace-augmented MUSIC with OSMP}
\acro{SCBSES}[SCBSES]{Source Compression Based Syndrome Encoding Scheme}
\acro{SCM}{sample covariance matrix}
\acro{SEP}{symbol error probability}
\acro{SG}[SG]{sparse-land Gaussian model}
\acro{SIMO}{single-input multiple-output}
\acro{SINR}{signal-to-interference plus noise ratio}
\acro{SIR}{signal-to-interference ratio}
\acro{SISO}{single-input single-output}
\acro{SMV}{single measurement vector}
\acro{SNR}[\textrm{SNR}]{signal-to-noise ratio} 
\acro{sp}[SP]{subspace pursuit}
\acro{SS}{spread spectrum}
\acro{SW}{sync word}
\acro{SAR}{specific absorption rate}
\acro{SSB}{synchronization signal block}
\acro{SDG}{Sustainable Development Goal}

\acro{TH}{time-hopping}
\acro{ToA}{time-of-arrival}
\acro{TR}{transmitted-reference}
\acro{TW}{Tracy-Widom}
\acro{TWDT}{TW Distribution Tail}
\acro{TCM}{trellis coded modulation}
\acro{TDD}{time-division duplexing}
\acro{TDMA}{Time Division Multiple Access}

\acro{UAV}{unmanned aerial vehicle}
\acro{uric}[URIC]{upper restricted isometry constant}
\acro{urict}[URICt]{upper restricted isometry constant threshold}
\acro{UWB}{ultrawide band}
\acro{UWBcap}[UWB]{Ultrawide band}   
\acro{URLLC}{Ultra Reliable Low Latency Communications}
         
\acro{wuric}[UWRIC]{upper weak restricted isometry constant}
\acro{wurict}[UWRICt]{UWRIC threshold}                
\acro{UE}{user equipment}
\acro{UL}{uplink}
\acro{ULA}{uniform linear array}

\acro{WiM}[WiM]{weigh-in-motion}
\acro{WLAN}{wireless local area network}

\acro{wm}[WM]{Wishart matrix}                               
\acroplural{wm}[WM]{Wishart matrices}
\acro{WMAN}{wireless metropolitan area network}
\acro{WPAN}{wireless personal area network}
\acro{wric}[WRIC]{weak restricted isometry constant}
\acro{wrict}[WRICt]{weak restricted isometry constant thresholds}
\acro{wrip}[WRIP]{weak restricted isometry property}
\acro{WSN}{wireless sensor network}                        
\acro{WSS}{wide-sense stationary}
\acro{WHO}{World Health Organization}
\acro{WP}{work package}

\acro{sss}[SpaSoSEnc]{sparse source syndrome encoding}
\acro{SO}{strategic objective}

\acro{VLC}{visible light communication}
\acro{RF}{radio frequency}
\acro{FSO}{free space optics}
\acro{IoST}{Internet of space things}

\acro{GSM}{Global System for Mobile Communications}
\acro{2G}{second-generation cellular networks}
\acro{3G}{third-generation cellular networsk}
\acro{4G}{fourth-generation cellular networks}
\acro{5G}{5th-generation cellular networks}	
\acro{6G}{6th-generation cellular networks}	
\acro{gNB}{next generation node B base station}
\acro{NR}{New Radio}
\acro{UN}{United Nations}
\acro{UMTS}{Universal Mobile Telecommunications Service}
\acro{LTE}{Long Term Evolution}

\acro{QoS}{Quality of Service}

\acro{ZZB}{Ziv-Zakai bound}

\acro{ESM}{explicit score matching}
\acro{ISM}{implicit score matching}
\acro{SSM}{sliced score matching}
\acro{DSM}{denoising score matching}
\acro{FSM}{Fisher score matching}
\acro{NCSN}{noise conditioned score network}
\acro{RCS}{radar cross section}
\acro{THz}{terahertz}
\acro{mmWave}{millimeter wave}
\end{acronym}

\begin{document}
\graphicspath{{./figures/}}
\title{
Scoring ISAC: Benchmarking Integrated Sensing and Communications via Score-Based \\ Generative Modeling
}
\author{
		Lin Chen, Chang Cai$^{\ast}$,  Huiyuan Yang, Xiaojun Yuan, Ying-Jun Angela Zhang
		\thanks{
        $^{\ast}$Corresponding author
        
               Lin Chen, Chang Cai, and Ying-Jun Angela Zhang are with the Department of Information Engineering, The Chinese University of Hong Kong (CUHK), Hong Kong (e-mail: \{cl022,~cc021,~yjzhang\}@ie.cuhk.edu.hk).
			
			Huiyuan Yang and Xiaojun Yuan are with the National Key Laboratory of Wireless Communications, University of Electronic Science and Technology of China (UESTC), Chengdu 611731, China (e-mail: hyyang@std.uestc.edu.cn; xjyuan@uestc.edu.cn).
		} 
	}

\maketitle


\begin{abstract}
Integrated sensing and communications (ISAC) is a key enabler for next-generation wireless systems, aiming to support both high-throughput communication and high-accuracy environmental sensing using shared spectrum and hardware. Theoretical performance metrics, such as mutual information (MI), minimum mean squared error (MMSE), and Bayesian Cram\'{e}r--Rao bound (BCRB), play a key role in evaluating ISAC system performance limits.
However, in practice, hardware impairments, multipath propagation, interference, and scene constraints often result in nonlinear, multimodal, and non-Gaussian distributions, making it challenging to derive these metrics analytically. Recently, there has been a growing interest in applying score-based generative models to characterize these metrics from data, although not discussed for ISAC. This paper provides a tutorial-style summary of recent advances in score-based performance evaluation, with a focus on ISAC systems. We refer to the summarized framework as {\em scoring ISAC}, which not only reflects the core methodology based on score functions but also emphasizes the goal of scoring (i.e., evaluating) ISAC systems under realistic conditions.
We present the connections between classical performance metrics and the score functions and provide the practical training techniques for learning score functions to estimate performance metrics. 
Proof-of-concept experiments on target detection and localization validate the accuracy of score-based performance estimators against ground-truth analytical expressions, illustrating their ability to replicate and extend traditional analyses in more complex, realistic settings.
This framework demonstrates the great potential of score-based generative models in ISAC performance analysis, algorithm design, and system optimization.
\end{abstract}


\section{Introduction}
\Ac{ISAC} has emerged as a transformative paradigm in next-generation wireless systems, enabling joint use of spectrum and hardware for both radio sensing and data communication~\cite{liu2022survey,lu2024integrated,liu2022integrated}. 
This paradigm shift is primarily driven by the growing demand for high-accuracy sensing in intelligent and autonomous applications such as autonomous driving~\cite{yurtsever2020survey}, \acp{UAV}~\cite{gupta2015survey}, \ac{IoT} networks~\cite{nguyen20216g}, and smart environments~\cite{ahmed2016internet}. These applications highly depend on real-time environmental awareness, including user localization, mobility tracking, and channel prediction, to support accurate perception and robust system operation. 
To meet such emerging requirements, \ac{6G} is expected to integrate sensing as a core service alongside communication~\cite{cui2021integrating,zhang2021overview}.
Meanwhile, modern sensing platforms benefit from the communication functionality to support distributed inference, cooperative sensing, and coordinated decision-making across spatially separated nodes~\cite{li2025multi}.
The integration of the sensing and communication functionalities is further facilitated by advances in ultra-high frequency (e.g., millimeter wave and terahertz)~\cite{wang2018millimeter,chen2019survey} and ultra-massive \ac{MIMO} technologies~\cite{faisal2020ultramassive}. 
These technologies offer high temporal and spatial resolution, enabling simultaneous high-throughput data transmission and precise environmental perception using the same signal and hardware~\cite{chen2022tutorial,demirhan2024cell,wang2023near}.
Motivated by the practical demands and technological advances, ISAC has attracted increasing research attention in both academia and industry~\cite{zhang2025integrated,3gpp2022isac,itu2021imt2030}. 


A key area of ISAC research lies in the analysis of its fundamental performance limits~\cite{liu2022survey,lu2024integrated,liu2022integrated,wang2024unified,wang2024cramer,xiong2023fundamental}. 
To characterize system capabilities from different but complementary perspectives, a wide range of metrics has been adopted from information theory, estimation theory, and detection theory.
Information-theoretic metrics, such as \ac{MI}, measure the maximum achievable data rate over a communication channel~\cite{shannon1948mathematical} and quantify the amount of information that a sensing system can extract from received signals about the environment or targets~\cite{wang2024unified}. 
Estimation-theoretic metrics, including \ac{CRB} and \ac{MMSE}, offer lower bounds on the variance of unbiased estimators and are commonly used in assessing the accuracy of sensing parameters, such as time delay, angle of arrival, and Doppler shift~\cite{boyer2010performance,chen2018impact}. 
Detection-theoretic metrics, such as detection probability, evaluate decision-making reliability in hypothesis testing, which is essential for tasks including target detection~\cite{ray2011radar} and symbol detection~\cite{sandberg1995some}. 

These theoretical metrics serve two fundamental roles in the development of ISAC systems.
First, they serve as performance benchmarks, revealing the gap between practical implementations and the fundamental limits. 
By quantifying this gap, researchers can better assess the effectiveness of existing algorithms and gain insights for improving system design to approach the theoretical potential~\cite{wang2024cramer,gurgunouglu2025performance,wang2023near}.  
Second, these metrics serve as optimization objectives, offering principled and quantifiable criteria for the design of waveforms, precoders, and resource allocation strategies that meet the dual requirements of sensing and communication. For instance, \ac{MI} or \ac{MMSE} can guide beamformer design for efficient data transmission~\cite{lin2019hybrid,el2014spatially}; CRB can be leveraged to optimize beamformers for high-resolution target localization~\cite{li2024near,liu2021cramer}. Moreover, \ac{MI} and \ac{MMSE} can be jointly optimized to design waveforms that support high communication and sensing performance~\cite{wang2024unified}.


While these theoretical metrics provide valuable benchmarks and design principles for developing ISAC systems, accurately computing these metrics in practical scenarios remains highly challenging.
A key difficulty stems from their strong dependence on the posterior distribution of unknown parameters (or channel inputs) given the observations (or channel outputs), which in turn is determined by the prior distributions, noise models, and likelihood functions.
In real-world ISAC scenarios, however, these distributions are often unknown and rarely available in closed form, rendering analytical evaluation of performance metrics intractable. 

To enable tractable analysis, many existing works adopt simplified statistical models~\cite{liu2022survey,lu2024integrated,liu2022integrated,wang2024unified,wang2024cramer,xiong2023fundamental}, e.g., assuming linear-Gaussian signal models or closed-form priors. 
While these assumptions lead to elegant formulations and closed-form expressions for theoretical metrics, they often fail to capture the statistical complexity of realistic sensing and communication environments.
For example, hardware impairments, such as power amplifier nonlinearity, IQ imbalance, phase noise, and ADC quantization~\cite{barneto2019full,koivunen2024multicarrier,wu2025quantized}, introduce distortions far from Gaussian distributed, which can dramatically alter channel capacity and error performance. 
Beyond hardware-induced effects, the signal propagation environment in both radar and communication systems is inherently complex and dynamic. Signals interact with clutter, interference, and multi-path propagation in ways that are often nonlinear and poorly approximated by linear-Gaussian models~\cite{de2021convergent,feintuch2023neural,haenggi2009interference}. 
In addition, sensing parameters such as target positions often lie on low-dimensional manifolds shaped by scene geometry~\cite{bae2022non}, e.g., movement constrained to walkable areas, deviating significantly from the assumptions of simple priors like isotropic Gaussian.
The mismatch between assumed and actual distributions may lead to misleading performance benchmarks.
Consequently, algorithms optimized under mismatched assumptions may perform suboptimally when deployed.

To overcome the limitations of classical model-based approaches (relying on hand-crafted statistical assumptions), deep generative modeling offers a promising data-driven alternative that directly learns the statistical distributions from complex real-world samples.
Among recent advances in deep generative models, score-based generative models~\cite{song2019generative,song2020score} (also known as diffusion models~\cite{ho2020denoising}) have emerged as a powerful tool for capturing high-dimensional, nonlinear, and multimodal data distributions.
These models learn the score function, i.e., the gradient of the log-density of the data distribution, which indicates the steepest ascent direction in the
log-density landscape. 
Once trained, the score guides sample generation by iteratively moving samples toward high-density regions through techniques such as Langevin dynamics~\cite{song2019generative,song2020score}, enabling the synthesis of diverse and highly realistic samples. 
With this capability, score-based generative models have set new benchmarks in fields such as image synthesis~\cite{dhariwal2021diffusion} and video generation~\cite{ho2022video}.
Beyond sample generation, score-based generative models have also been successfully applied to inverse problems~\cite{song2023pseudoinverse,chung2023decomposed,daras2024survey} and image reconstruction tasks~\cite{wang2022zero,zhu2023denoising}, e.g., magnetic resonance imaging (MRI) and computed tomography (CT).
In these applications, the score-based priors have been used to replace traditional hand-crafted priors. These learned prior scores better capture the nonlinear geometry and multimodal nature of real-world data spaces, enabling more accurate reconstructions.

Recent studies have shown that performance metrics, such as \ac{MI}, \ac{MMSE}, and \ac{BCRB}, can be reformulated in terms of score functions~\cite{franzese2023minde,kong2023information,kong2023interpretable,cai2025score, cai2025massive_connectivity, xue2024score, wang2025traversing,crafts2024bayesian,habi2025learned}.
Thus, score-based generative models open up new possibilities for estimating these metrics in a data-driven manner.  
In essence, the evaluation of performance metrics typically requires access to the underlying distribution of random variables and their conditional distributions, which are often analytically intractable in real-world scenarios. Score-based generative models address this challenge by learning the score functions of these distributions directly from data, enabling accurate estimation under the score-based reformulation of these performance metrics. This data-driven approach offers a powerful alternative to the classical model-based approach, which relies on idealized assumptions that may not hold in practice.
 
Despite their growing impact in other domains, score-based generative modeling has so far received very limited research attention in ISAC. To bridge this gap, this paper provides a tutorial-style summary of recent advances in score-based performance evaluation, with a particular focus on ISAC systems. We present the scoring ISAC framework, a data-driven framework using score-based generative modeling for characterizing the ISAC system performance limits. Here, the term {\em scoring ISAC} reflects both the supporting methodology: using {\em score functions to capture underlying distributions in ISAC systems}, and the ultimate goal of scoring ISAC: {\em benchmarking the performance of ISAC systems}.  
Specifically, we present the explicit connections between score functions and ISAC performance metrics from information, estimation, and detection theory. We also present the practical training techniques for learning the score functions required in performance evaluation.
To validate the framework, we conduct proof-of-concept experiments, including target detection and localization tasks. The results show that score-based estimators can closely match analytical benchmarks while offering greater capability in complex, non-ideal environments. 
Together, the scoring ISAC framework highlights the great potential of score-based generative modeling in performance analysis, algorithm design, and system optimization for future ISAC development.


\begin{figure*}
    \centering
    \includegraphics[width=0.85\linewidth]{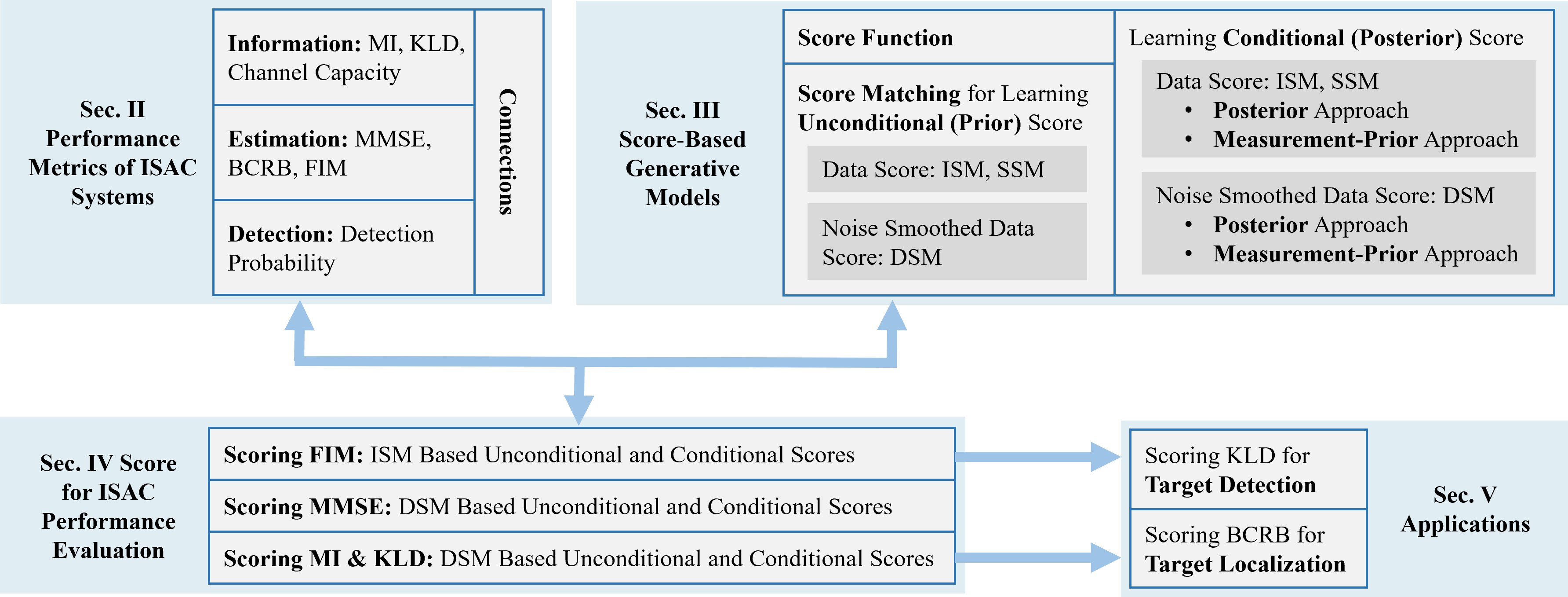}
    \caption{Paper structure.}
    \label{fig:Paper}
\end{figure*}

\begin{figure*}
    \centering
    \includegraphics[width=0.65\linewidth]{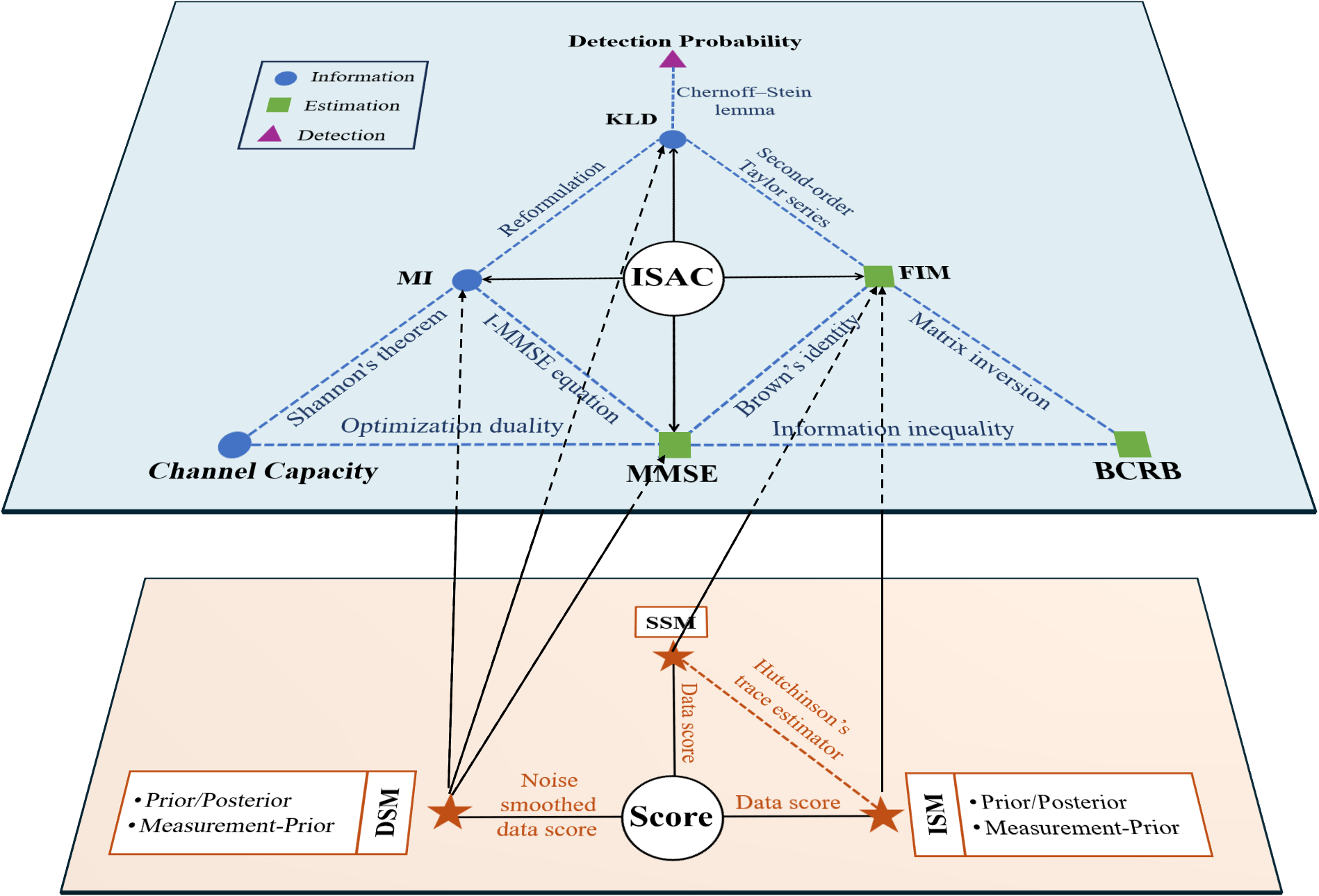}
    \caption{Framework of Scoring ISAC. The bottom layer provides the foundational technologies for evaluating the performance metrics shown in the top layer.}
    \label{fig:Framework}
\end{figure*}

{\em Organization:}
The remainder of this paper is organized as shown in Fig.~\ref{fig:Paper}. Sec.~\ref{sec:ISACmetrics} provides the fundamental metrics of \ac{ISAC} systems from information, estimation, and detection theories and connections between metrics, as shown in the top layer in Fig.~\ref{fig:Framework}.
Sec.~\ref{sec:ScoreBasis} introduces the fundamentals of score-based generative models, as shown in the ground layer in Fig.~\ref{fig:Framework}.
Sec.~\ref{sec:Score-ISAC} presents the methods to benchmark ISAC via score-based generative
modeling. The applications are given in Sec.~\ref{sec:Sapplication}. Finally, Sec.~\ref{sec:conclusion} concludes this paper and discusses future extensions.

{\em Notations}: 
		The upper- and lower-case bold letters denote matrices and vectors, respectively. 
		For a matrix, $\mathbf{(\cdot)}^\mathsf{T}$, $\mathbf{(\cdot)}^{*}$, and $\mathbf{(\cdot)}^\mathsf{H}$, $\mathbf{(\cdot)}^{-1}$ represent transpose, conjugate, conjugate transpose, and inverse operators, respectively. $\left\|\cdot\right\|_2$ denote the $l_2$ norm. $\mathbf{I}$ denotes the identity matrix of an appropriate size. $\mathcal{N}(\mathbf{0},\mathbf{I})$ denotes a standard Gaussian distribution, and $\mathcal{CN}(\mathbf{0},\mathbf{I})$ denotes a standard complex Gaussian distribution.
For a scalar-valued function $f:\mathbb{R}^{D}\rightarrow\mathbb{R}$, its gradient at $\mathbf{x}\in\mathbb{R}^{D}$ is a column vector $\nabla_{\mathbf{x}} f(\mathbf{x})=\left[\frac{\partial f}{\partial [\mathbf{x}]_{1}},..., \frac{\partial f}{\partial [\mathbf{x}]_{d}},...,\frac{\partial f}{\partial [\mathbf{x}]_{D}}\right]^\mathsf{T}$, where $[\mathbf{x}]_{d}$ is the $d$-th element of $\mathbf{x}$.
For a vector-valued function $\bm{f}:\mathbb{R}^{D_1}\rightarrow\mathbb{R}^{D_2}$, its Jacobian at $\mathbf{x}\in \mathbb{R}^{D_1}$  is a $D_2\times D_1 $ matrix $\nabla_{\mathbf{x}} \bm{f}(\mathbf{x})$ with the $(d_2,d_1)$-th element being $[\nabla_{\mathbf{x}} \bm{f}(\mathbf{x})]_{d_2,d_1}=\frac{\partial [\bm{f}(\mathbf{x})]_{d_2}}{\partial [\mathbf{x}]_{d_1}}$. 
For random vectors $\mathbf{x}$ and $\mathbf{y}$, $p(\mathbf{x})$ represents the \ac{PDF} of $\mathbf{x}$, and $\mathbb{E}_{p(\mathbf{x})}[\cdot]$ represents the expectation with respect to $\mathbf{x}$.
The dataset $\{\mathbf{x}_1,..., \mathbf{x}_M\}$ is abbreviated as $\mathcal{X}_M=\{\mathbf{x}_m\}_{m=1}^{M}$.

\section{Performance Metrics of ISAC Systems}\label{sec:ISACmetrics}

This section introduces key performance metrics used to characterize the performance of \ac{ISAC} systems.
We then examine the intrinsic connections between these metrics.  

\subsection{Information Theoretic Metrics}\label{sec:information}
Consider an \ac{ISAC} system consisting of a full-digital \ac{BS}, a communication user, and a sensing target~\cite{wang2024unified}. The \ac{BS} transmits the signal $\mathbf{x}$ to simultaneously communicate with the user and sense the target. The communication user receives the signal $\mathbf{y}_{\rm c}$, which is given by
\begin{align}\label{eq:yc}        
		\mathbf{y}_{\rm c}&=\mathbf{H}_{\rm c}\mathbf{x}+\mathbf{z}_{\rm c},
\end{align}
where $\mathbf{H}_{\rm c}$ is the communication channel (from the \ac{BS} to the user), and $\mathbf{z}_{\rm c}$ represents the noise at the user.
Meanwhile, the sensing target reflects $\mathbf{x}$ and then the BS receives the echo $\mathbf{y}_{\rm s}$, which is given by
\begin{align}\label{eq:ys}
\mathbf{y}_{\rm s}&=\mathbf{H}_{\rm s}\mathbf{x}+\mathbf{z}_{\rm s},
	\end{align}   
where $\mathbf{H}_{\rm s}$ denotes the round-trip sensing channel (from the \ac{BS} to the target and back to the \ac{BS}), and $\mathbf{z}_{\rm s}$ represents the noise at the \ac{BS}.

\subsubsection{Communication MI}
In the communication scenario, the channel $\mathbf{H}_{\rm c}$ is typically obtained via channel estimation methods.
The \ac{MI} between the transmitted signal $\mathbf{x}$ and the received signal $\mathbf{y}_{\rm c}$, conditioned on the communication channel $\mathbf{H}_{\rm c}$, quantifies the amount of information the user can infer about $\mathbf{x}$ from $\mathbf{y}_{\rm c}$.
This quantity, referred to as the communication \ac{MI}, is defined as~\cite{goldsmith2005wireless}
\begin{align}\label{eq:CMI}
	\begin{split}
		I(\mathbf{x},\mathbf{y}_{\rm c}|\mathbf{H}_{\rm c})&=\mathbb{E}_{p(\mathbf{x},\mathbf{y}_{\rm c},\mathbf{H}_{\rm c})}\left[\log \frac{p(\mathbf{x}|\mathbf{y}_{\rm c},\mathbf{H}_{\rm c})}{ p(\mathbf{x}|\mathbf{H}_{\rm c})}\right]. 
	\end{split}
\end{align}
Assuming that the transmitted signal $\mathbf{x}$ is independent of the communication channel $\mathbf{H}_{\rm c}$, we have $p(\mathbf{x}|\mathbf{H}_{\rm c})= p(\mathbf{x})$.
Eq.~\eqref{eq:CMI} can be rewritten in terms of the \ac{KLD} as~\cite{kailath2003divergence}
\begin{align}  
	\begin{split}
		I(\mathbf{x},\mathbf{y}_{\rm c}|\mathbf{H}_{\rm c})=\mathbb{E}_{p(\mathbf{y}_{\rm c},\mathbf{H}_{\rm c})} [D_{\rm KL}\left(p(\mathbf{x}|\mathbf{y}_{\rm c},\mathbf{H}_{\rm c})||  p(\mathbf{x})\right)],   
	\end{split}
\end{align}
where $D_{\rm KL}(p|| q)$ is the \ac{KLD} between the distributions $p(\cdot)$ and $q(\cdot)$.
According to Shannon's theorem~\cite{shannon1948mathematical}, the channel capacity is the maximum achievable \ac{MI} over all possible input distributions:
\begin{align}
	C=\max_{p(\mathbf{x})} I(\mathbf{x},\mathbf{y}_{\rm c}|\mathbf{H}_{\rm c}).
\end{align}

\subsubsection{Sensing MI}
In contrast to the communication setting, the sensing receiver, i.e., BS, knows the transmitted signal $\mathbf{x}$, and the sensing goal is to extract information about the sensing channel $\mathbf{H}_{\rm s}$ from the received sensing echo $\mathbf{y}_{\rm s}$.
The \ac{MI} between the sensing channel $\mathbf{H}_{\rm s}$ and the received echo $\mathbf{y}_{\rm s}$, conditioned on the transmitted signal $\mathbf{x}$, is defined as the sensing \ac{MI}~\cite{wang2024unified,SMI}:
\begin{align}  \label{eq:SMI}
	\begin{split}
		I(\mathbf{H}_{\rm s},\mathbf{y}_{\rm s}|\mathbf{x})&=\mathbb{E}_{p(\mathbf{H}_{\rm s},\mathbf{y}_{\rm s},\mathbf{x})}\left[\log \frac{p(\mathbf{H}_{\rm s}|\mathbf{y}_{\rm s},\mathbf{x})}{ p(\mathbf{H}_{\rm s}|\mathbf{x})}\right]
		\\& = \mathbb{E}_{p(\mathbf{y}_{\rm s},\mathbf{x})} [D_{\rm KL}\left(p(\mathbf{H}_{\rm s}|\mathbf{y}_{\rm s},\mathbf{x})||  p(\mathbf{H}_{\rm s})\right)],  
	\end{split}
\end{align}
where the equivalence $p(\mathbf{H}_{\rm s}|\mathbf{x}) = p(\mathbf{H}_{\rm s})$ follows from the assumption that $\mathbf{x}$ and $\mathbf{H}_{\rm s}$ are independent.
The sensing MI quantifies the expected information gain about the unknown sensing environment $\mathbf{H}_{\rm s}$ from the received signal $\mathbf{y}_{\rm s}$, under a known transmitted signal $\mathbf{x}$.
It provides a principled metric for evaluating the quality of target estimation or environmental reconstruction in \ac{ISAC} systems.

\subsection{Estimation Theoretic Metrics}\label{sec:estimation}
Estimation is the process of extracting unknown parameters of interest from observed data (or measurements). In \ac{ISAC} systems, these parameters include \ac{CSI}, transmitted signals, or target locations.
A general parameter estimation model can be formulated as
\begin{align}\label{eq:estimation}
    \mathbf{y}={\bm f}(\bm{\theta})+\mathbf{z},
\end{align}
where $\bm{\theta}$ is the unknown parameter to be estimated from the measurement $\mathbf{y}$, ${\bm f}(\cdot)$ is the measurement operator determined by the \ac{ISAC} system, and $\mathbf{z}$ is the measurement noise. 
Several metrics, e.g., \ac{MMSE}, \ac{CRB}, and \ac{FIM}, are commonly used to evaluate the estimation performance~\cite{kay1993fundamentals}.

\subsubsection{\ac{MMSE}}
Denote $\hat{\bm{\theta}}$ as an estimate of $\bm{\theta}$. 
The optimal estimator that minimizes the \ac{MSE} is given by the posterior mean~\cite{kay1993fundamentals}, i.e.,
\begin{align}\label{eq:mmse_mean}
    \hat{\bm{\theta}}_{\rm mmse}=\arg \min_{\hat{\bm{\theta}}} \mathbb{E}_{p(\bm{\theta},\mathbf{y})}\left[\left\|\bm{\theta}-\hat{\bm{\theta}}\right\| ^2_2\right]=\mathbb{E}_{p(\bm{\theta}|\mathbf{y})} [\bm{\theta}].
\end{align}
The corresponding \ac{MMSE} is given by
\begin{align}\label{eq:mmse_theta}
    {\rm MMSE}=\mathbb{E}_{p(\bm{\theta},\mathbf{y})}\left[\left\| \bm{\theta}-\hat{\bm{\theta}}_{\rm mmse} \right\|  ^2_2\right].
\end{align}

\subsubsection{\ac{CRB} and \ac{FIM}}\label{subsubsec:CRB-FIM}
In a classical parameter estimation problem, $\bm\theta$ is treated as a deterministic but unknown constant, and \ac{CRB} provides a lower bound on the \ac{MSE} of any unbiased estimator~\cite{rao1992information}. \ac{BCRB} extends the concept of \ac{CRB} to the Bayesian framework, where $\bm \theta$ is treated as a random variable with a known prior distribution $p(\bm \theta)$~\cite{trees2007bayesian}.
The \ac{MSE} of $\hat{\bm{\theta}}$ satisfies the following information inequality~\cite{van2004detection}
\begin{align}\label{eq:BFIM}
\mathbb{E}_{p(\bm{\theta},\mathbf{y})}\left[(\bm{\theta}-\hat{\bm{\theta}})(\bm{\theta}-\hat{\bm{\theta}})^\mathsf{T}\right] \succeq \mathbf{V}_{\rm b} \overset{\triangle}{=} \mathbf{J}_{\rm b}^{-1},
\end{align}
where $\mathbf{J}_{\rm b}\overset{\triangle}{=}\mathbb{E}_{p(\bm \theta,\mathbf{y})}[\nabla_{\bm \theta}\log p(\bm \theta,\mathbf{y}) \nabla_{\bm \theta}\log p(\bm \theta,\mathbf{y})^\mathsf{T}]$ is the Bayesian \ac{FIM} and  $\mathbf{V}_{\rm b}$ is the \ac{BCRB} matrix.
The scalar form of the \ac{BCRB} is ${\rm BCRB}={\rm Tr}(\mathbf{V}_{\rm b})={\rm Tr}(\mathbf{J}_{\rm b}^{-1})$, which bounds the \ac{MSE} as 
\begin{align}
\mathbb{E}_{p(\bm{\theta},\mathbf{y})}\left[\left\| \bm{\theta}-\hat{\bm{\theta}} \right\| ^2_2\right]\ge {\rm MMSE}\ge {\rm BCRB}.
\end{align}

The Bayesian \ac{FIM} $\mathbf{J}_{\rm b}$ can also be expressed in terms of the posterior distribution~\cite{van2004detection}. Specifically, we know from probability theory that the joint probability can be decomposed as the product of the posterior probability and the marginal probability of the data, i.e., $p(\bm \theta,\mathbf{y}) =p(\bm \theta|\mathbf{y})p(\mathbf{y})$. Moreover, the gradient of $\log p(\mathbf{y})$ with respect to $\bm \theta$ is zero, i.e, $\nabla_{\bm \theta}\log p(\mathbf{y})=\mathbf{0}$. Therefore, the posterior based formulation of Bayesian \ac{FIM} is~\cite{van2004detection}
\begin{align}\label{eq:BFIM-posterior}
    \begin{split}
        \mathbf{J}_{\rm b}=\mathbb{E}_{p(\mathbf{y},\bm{\theta})}[\nabla_{\bm \theta}\log p(\bm \theta|\mathbf{y}) \nabla_{\bm \theta}\log p(\bm \theta|\mathbf{y})^\mathsf{T}].
    \end{split}
\end{align}

Another way of calculating the Bayesian \ac{FIM} is based on the measurement distribution $p(\mathbf{y}|\bm \theta)$ and the prior distribution $p(\bm \theta)$.
Specifically, with $p(\bm \theta,\mathbf{y}) =p(\mathbf{y}|\bm \theta)p(\bm \theta)$, the Bayesian \ac{FIM} can be reformulated as~\cite{van2004detection,crafts2024bayesian}
\begin{align}\label{eq:BFIM-Bayes}
    \mathbf{J}_{\rm b} =  \mathbf{J}_{\rm p}+\mathbf{J}_{\rm d}.
\end{align}
Here, $\mathbf{J}_{\rm p}$ is the prior \ac{FIM} provided by the prior distribution $p(\bm{\theta})$, which is given by 
\begin{align}\label{eq:priorFIM}
    \mathbf{J}_{\rm p}&\overset{\triangle}{=}\mathbb{E}_{p(\bm{\theta})}[\nabla_{\bm \theta}\log p(\bm \theta) \nabla_{\bm \theta}\log p(\bm \theta)^\mathsf{T}],
\end{align}
and $\mathbf{J}_{\rm d}$ is the data FIM (i.e., the average Fisher information associated with the observed data) which is given by
    \begin{align}\label{eq:FIM_D}
        \mathbf{J}_{\rm d}&=\mathbb{E}_{p(\mathbf{y},\bm{\theta})}[\nabla_{\bm \theta}\log p(\mathbf{y}|\bm \theta) \nabla_{\bm \theta}\log p(\mathbf{y}|\bm \theta)^\mathsf{T}]\nonumber\\&= \mathbb{E}_{p(\bm{\theta})}\big[\,  \underbrace{\mathbb{E}_{p(\mathbf{y}|\bm \theta)}[ \nabla_{\bm \theta}\log p(\mathbf{y}|\bm \theta) \nabla_{\bm \theta}\log p(\mathbf{y}|\bm \theta)^\mathsf{T}]}_{\mathbf{J}(\bm \theta)}\,\big],
\end{align}
where $\mathbf{J}(\bm \theta)$ represents the classical \ac{FIM}.
Specially, for a deterministic $\bm \theta$, the \ac{CRB} matrix is given by $\mathbf{J}(\bm \theta)^{-1}$~\cite{kay1993fundamentals}.

\subsection{Detection Theoretic Metrics}\label{sec:detection}

Detection involves making decisions about which hypothesis is true based on noisy measurements, from a predefined set of binary or multiple hypotheses. In \ac{ISAC} systems, detection tasks include demodulating communication symbols over discrete constellations, identifying the presence or absence of a target, and distinguishing between communication signals and sensing echoes.
To illustrate the detection process, we take a target detection problem as an example.
A full-digital BS transmits the signal $\mathbf{x}$ repeatedly over $K$ sensing intervals to obtain i.i.d. measurements.
We denote the binary hypotheses regarding the presence or absence of the target as $\mathcal{H}_1$ or $\mathcal{H}_0$, respectively.
For simplicity, we assume a clutter-free environment, where clutter echoes are effectively suppressed using standard techniques such as constant false alarm rate (CFAR) detection~\cite{clutter-free,wang2024cramer}. 
The $k$-th measurement under each hypothesis can be formulated as 
\begin{align}
    \begin{cases}
        \mathcal{H}_1: \mathbf{y}_{\rm s}[k]=\mathbf{H}_{\rm s}\mathbf{x}+\mathbf{z}_{\rm s}[k],\\
        \mathcal{H}_0: \mathbf{y}_{\rm s}[k]=\mathbf{z}_{\rm s}[k],\\
    \end{cases}
\end{align}
where $\mathbf{H}_{\rm s}$ and $\mathbf{z}_{\rm s}[k]$ are defined in \eqref{eq:ys}, and $k\in\{1,..., K\}$.
The statistical behavior of $\mathbf{y}_{\rm s}[k]$ under each hypothesis is described by the likelihoods $p_0(\mathbf{y}_{\rm s}[k])=p(\mathbf{y}_{\rm s}[k]|\mathcal{H}_0)$ and $p_1(\mathbf{y}_{\rm s}[k])=p(\mathbf{y}_{\rm s}[k]|\mathcal{H}_1)$, respectively.  
Based on the $K$ observations $\{\mathbf{y}_{\rm s}\}_{k=1}^K$, the detector infers which hypothesis is more likely.

For binary detection, there are two types of error probabilities: the probability of false
alarm (type \uppercase\expandafter{\romannumeral1} error), denoted by $P_{\rm fa}=\mathbb{P}\{\text{decide } \mathcal{H}_1|\mathcal{H}_0\}$, and the probability of miss detection (type \uppercase\expandafter{\romannumeral2} error), denoted by $P_{\rm md}=\mathbb{P}\{\text{decide } \mathcal{H}_0|\mathcal{H}_1\}$. Correspondingly, the detection probability, i.e., the probability that the system
correctly identifies a target when it is present, is $P_{\rm d}=\mathbb{P}\{\text{decide } \mathcal{H}_1|\mathcal{H}_1\}=1-P_{\rm md}$.
A classical decision rule for binary detection is the likelihood ratio test (LRT):
\begin{align}\label{eq:LRT}
    \Lambda=\prod_{k=1}^{K}\frac{p_1(\mathbf{y}_{\rm s}[k])}{p_0(\mathbf{y}_{\rm s}[k])}  \mathop{\gtrless}_{\mathcal{H}_0}^{\mathcal{H}_1} \eta ,
\end{align}
where $\eta$ is a predetermined threshold that controls the trade-off between $P_{\rm fa}$ and $P_{\rm d}$.
To make this concrete, suppose $p_0(\mathbf{y}_{\rm s}[k])=\mathcal{CN}(\mathbf{0}, \sigma^2_{\rm s}\mathbf{I})$ and $p_1(\mathbf{y}_{\rm s}[k])=\mathcal{CN}(\mathbf{\mathbf{H}_{\rm s}\mathbf{x}}, \sigma^2_{\rm s}\mathbf{I})$ are Gaussian distributions. Then, the LRT gives 
\begin{align}\label{eq:Pfa}
    P_{\rm fa}&=\mathbb{P}\{\Lambda>\eta\,|\,\mathcal{H}_0\}=  Q\left( \frac{\eta'+K\gamma}{\sqrt{2 K \gamma }} \right),
    \\P_{\rm d}&=\mathbb{P}\{\Lambda>\eta\,|\,\mathcal{H}_1\}=Q\left( \frac{\eta' - K\gamma }{\sqrt{2 K \gamma }} \right),\label{eq:Pd}
\end{align} 
where $Q(\cdot)$ denotes the standard Gaussian tail function, $\eta'=\log\eta$, and $\gamma=\frac{\left\|\mathbf{H}_{\rm s}\mathbf{x}\right\|_2^2}{\sigma_{\rm s}^2}$ is the \ac{SNR}.
Eqs.~\eqref{eq:Pfa}-\eqref{eq:Pd} reveal the inherent trade-off: increasing $\eta$ decreases both $P_{\rm fa}$ and $P_{\rm d}$. In practical systems, $P_{\rm fa}$ is often fixed to meet a specific system constraint, and the objective becomes maximizing $P_{\rm d}$, or equivalently, minimizing $P_{\rm md}$, under this constraint. 
Specifically,  by setting an appropriate threshold $\eta$ based on \eqref{eq:Pfa} to fix $P_{\rm fa}$ under the LRT, and substituting into \eqref{eq:Pd}, $P_{\rm d}$ can be expressed as~\cite{cover1999elements}
\begin{align}\label{eq:pd_pfa}
P_{\rm d} = Q\left( Q^{-1}(P_{\rm fa}) - \sqrt{ \frac{2K \left\|\mathbf{H}_{\rm s}\mathbf{x}\right\|_2^2 }{\sigma_{\rm s}^2 } } \right).
\end{align}
This highlights that under a fixed false alarm constraint, the detection probability improves as \ac{SNR} $\frac{\left\|\mathbf{H}_{\rm s}\mathbf{x}\right\|_2^2}{\sigma_{\rm s}^2}$ or the number of observations $K$ increases.

From the LRT in \eqref{eq:LRT}, the detection performance fundamentally depends on the statistical distinguishability of the distributions under the two hypotheses. Intuitively, if $p_1(\mathbf{y}_{\rm s}[k])$ and $p_0(\mathbf{y}_{\rm s}[k])$ are identical, the two hypotheses are statistically indistinguishable, making reliable detection impossible. Conversely, increasing the divergence between these distributions enables more accurate target identification. This divergence can be measured by the \ac{KLD} as 
\begin{align}
    D_{\rm KL}\left( p_0  ||  p_1 \right)
\!=\mathbb{E}_{ p_0(\mathbf{y}_{\rm s}[k])}[\log p_0(\mathbf{y}_{\rm s}[k]) -\log p_1(\mathbf{y}_{\rm s}[k])].
\end{align}
In the Gaussian example considered above, the KLD is
\begin{align}\label{eq:KLDgaussian}
D_{\rm KL}\left( p_0  ||  p_1 \right)=\frac{\left\|\mathbf{H}_{\rm s}\mathbf{x}\right\|_2^2}{\sigma_{\rm s}^2}.
\end{align}
Substituting \eqref{eq:KLDgaussian} into \eqref{eq:pd_pfa}, the miss detection probability and the KLD are connected via
\begin{align}\label{eq:Pmd}
    &P_{\rm md} = 1 - P_{\rm d}  =Q\left( \sqrt{ 2K D_{\rm KL}(p_0||p_1) }-Q^{-1}(P_{\rm fa}) \right)
    \nonumber\\&\overset{(a)}{\approx}
    \frac{1}{2}\exp\left(-\frac{\left( \sqrt{2K D_{\rm KL}(p_0||p_1)} - Q^{-1}(P_{\rm fa})\right)^2}{2}\right),
\end{align}
where (a) is from $Q(x)\approx \frac{1}{2}\exp\left(-\frac{x^2}{2}\right)$ for large $x>0$~\cite{abramowitz1948handbook}, which holds when the number of observations $K$ is sufficiently large.  From~\eqref{eq:Pmd}, the dominant term in the exponent scales linearly with $K$, implying an exponential decay in $P_{\rm md}$ or, equivalently, exponential growth in $P_{\rm d}$, as $K$ increases. This exponential behavior is formally established by the Chernoff-Stein lemma~\cite{cover1999elements}:
\begin{align}\label{eq:KLD-target}
  \lim_{K\to\infty} \left( -\frac{1}{K} \log (1-P_{\rm d})\right)=D_{\rm KL}\left( p_0  ||  p_1 \right).
\end{align}
Therefore, the KLD not only serves as a measure of statistical dissimilarity between hypotheses but also determines the asymptotic rate at which the detection probability increases. In this sense, improving detection reliability in \ac{ISAC} systems can be systematically approached by maximizing the KLD between hypotheses.

\subsection{Connections Between Metrics}

The connections between the aforementioned ISAC metrics are visually presented in the top layer of Fig.~\ref{fig:Framework}, with a detailed explanation provided below.

\subsubsection{MI-MMSE} \label{subsubsec:MI-MMSE}
Consider an \ac{AWGN} channel model:
\begin{align}\label{eq:xt}
    \mathbf{x}_t=\mathbf{x} + \sigma_t \mathbf{n},
\end{align}
where $\mathbf{x}\in\mathbb{R}^{D}$ is the channel input, $\mathbf{n}\sim \mathcal{N}(\mathbf{0}, \mathbf{I})$ is the standard Gaussian noise, ${\rm snr}=\sigma_t^{-2}$ represents the \ac{SNR}, $\sigma_t\ge0$, and $\mathbf{x}_t$ is the noisy output.
For a given \ac{SNR}, the \ac{MMSE} in estimating $\mathbf{x}$ from $\mathbf{x}_t$ can be expressed as
\begin{align}
    {\rm MMSE}(\sigma_t)=\mathbb{E}_{p(\mathbf{x},\mathbf{x}_t)}\left[ \left\| \mathbf{x}-\hat{\mathbf{x}}_{\rm mmse}(\mathbf{x}_t,\sigma_t) \right\|   _2^2\right],
\end{align}
where $\hat{\mathbf{x}}_{\rm mmse}(\mathbf{x}_t,\sigma_t)=\mathbb{E}_{p(\mathbf{x}|\mathbf{x}_t)} [\mathbf{x}]$ is the \ac{MMSE} estimator~\cite{I-MMSE_FIM-MI1}. 
Denote the \ac{MI} between $\mathbf{x}$ and $\mathbf{x}_t$ as $I(\mathbf{x}, \mathbf{x}_t)$.
The I-MMSE equation relates \ac{MMSE} to \ac{MI} by~\cite{I-MMSE_FIM-MI1,I-MMSE_FIM-MI2}
\begin{align}\label{eq:I-MMSE}
    \frac{\mathrm{d} I(\mathbf{x}, \mathbf{x}_t)}{\mathrm{d}{\,\rm snr}} & = \frac{1}{2}{\rm MMSE}({\sigma_t}).
\end{align}
Eq.~\eqref{eq:I-MMSE} reveals that in an AWGN channel, the rate at which \ac{MI} grows with \ac{SNR} is exactly half of the \ac{MMSE}. 
Intuitively, the better we can estimate a signal from its noisy observation (i.e., lower \ac{MMSE}), the more efficiently we can leverage the increased \ac{SNR} to transmit information (i.e., higher growth rate of \ac{MI}).

\subsubsection{MI-FIM}
The connection between MI and FIM is established via De-Bruijn's identity~\cite{FIM-MI1,FIM-MI2}. Specifically, for the same \ac{AWGN} channel model in \eqref{eq:xt}, De-Bruijn's identity relates the differential entropy of $\mathbf{x}_t$ to its \ac{FIM} by~\cite{I-MMSE_FIM-MI1,I-MMSE_FIM-MI2,FIM-MI1,FIM-MI2}
\begin{align}\label{eq:H-FIM0}
    \frac{\mathrm{d} h( \mathbf{x}_t)}{\mathrm{d}\sigma_t^2} & = \frac{1}{2} {\rm Tr}\left(\mathbf{J}(\mathbf{x}_t)\right),
\end{align}
where $h( \mathbf{x}_t)$ is the differential entropy of $\mathbf{x}_t$ and $ \mathbf{J}(\mathbf{x}_t)=\mathbb{E}_{p(\mathbf{x}_t)}\left[\nabla_{\mathbf{x}_t}\log p(\mathbf{x}_t) \nabla_{\mathbf{x}_t}\log p(\mathbf{x}_t)^\mathsf{T}\right]$ is the FIM of $\mathbf{x}_t$. 
Meanwhile, the \ac{MI} $I(\mathbf{x}, \mathbf{x}_t)$ can be expressed in terms of entropy as~\cite{cover1999elements}
\begin{align}\label{eq:MI-H}
I(\mathbf{x},\mathbf{x}_t)=h(\mathbf{x}_t) -h(\mathbf{x}_t| \mathbf{x} ) = h(\mathbf{x}_t) -\frac{D}{2}\log (2\pi e \sigma_t^2),
\end{align}
where $h(\mathbf{x}_t| \mathbf{x} )$ is the conditional differential entropy, which equals the entropy of the Gaussian noise $\sigma_t\mathbf{n}\in\mathbb{R}^D$.
Substituting \eqref{eq:MI-H} into \eqref{eq:H-FIM0} yields the \ac{MI}-\ac{FIM} relationship:
\begin{align}\label{eq:MI-FIM}
    \frac{\mathrm{d} I( \mathbf{x},\mathbf{x}_t)}{\mathrm{d}\,{\rm snr}} & = \frac{D}{2 {\rm snr}}-\frac{1}{2 {\rm snr}}{\rm Tr}\left(\mathbf{J}(\mathbf{x}_t)\right).
\end{align}

\subsubsection{MMSE-FIM}
Combining the MI-MMSE relation in \eqref{eq:I-MMSE} with the MI-FIM relation in \eqref{eq:MI-FIM}, for the same \ac{AWGN} channel model in \eqref{eq:xt}, the connection between MMSE and FIM is given by~\cite{I-MMSE_FIM-MI1}
\begin{align}\label{eq:MMSE-FIM}
    \sigma_t^{2} {\rm Tr}\left(\mathbf{J}(\mathbf{x} + \sigma_t \mathbf{n})\right)=D-\frac{1}{\sigma_t^{2}}{\rm MMSE}({\sigma}_t).
\end{align}
For the scalar AWGN channel with $D=1$, the relation \eqref{eq:MMSE-FIM} simplifies to a well-known Brown's identity~\cite{MMSE-FIM,brown1971admissible}, i.e., for $x\in\mathbb{R}$ and $n\sim \mathcal{N}(0,1)$,
\begin{align}\label{eq:brown}
    {\rm J}(\sqrt{{\rm snr}}\,x + n)=1-{\rm snr}\,{\rm MMSE}({\sigma}_t), 
\end{align}
where ${\rm J}(\cdot)$ is the Fisher information, ${\rm J}(x + \sigma_t n)={\rm snr}\,{\rm J}(\sqrt{{\rm snr}}\,x+n)$~\cite{I-MMSE_FIM-MI1}, and ${\rm snr}=\sigma_t^{-2}$.
The identity \eqref{eq:brown} highlights the inverse relationship between estimation error and Fisher information, i.e., as the estimation error decreases, the Fisher information of the input contained in the noisy observation increases. 

\subsubsection{KLD-FIM}
Fisher information characterizes the curvature of \ac{KLD} between likelihoods. Specifically, for the general parameter estimation model in \eqref{eq:estimation}, $\mathbf{J}(\bm \theta)$ represents the observed \ac{FIM}, as defined in \eqref{eq:FIM_D}, which measures how sensitively the likelihood function responds to small changes in the parameter $\bm \theta$.
For a small perturbation $ {\bm\delta}$ around the true parameter $\bm{\theta}$, the \ac{KLD} between likelihoods $p(\mathbf{y}|\bm \theta)$ and $p(\mathbf{y}|\bm \theta+{\bm\delta})$ can be approximated by expanding $\log p(\mathbf{y}|\bm \theta+{\bm\delta})$ in a second-order Taylor series expansion around $\bm \theta$, i.e.,~\cite{kullback1997information,FIM-KLD}
\begin{align}\label{eq:KLD-FIM}
\begin{split}
&D_{\text{KL}}\left(\,p(\mathbf{y}|\bm \theta) \,||\, p(\mathbf{y}|\bm \theta+{\bm\delta}) \,\right) \approx
 {\bm\delta}^\mathsf{T} \mathbf{J}(\bm \theta)
{\bm\delta}.
\end{split}
\end{align}
Intuitively, a larger Fisher information implies that the likelihood function is more sharply curved around the true parameter $\bm \theta$, meaning that small perturbations in the parameter will lead to a larger increase in the KLD.

\subsubsection{Capacity-MMSE}
Channel capacity maximization and MMSE minimization problems are closely related through optimization duality.
Recent studies have demonstrated an equivalence between the weighted sum-rate maximization problem (corresponding to channel capacity) and the matrix-weighted sum-MSE minimization problem (related to MMSE)~\cite{MMSE-Capacity, christensen2008weighted, shen2018fractional}. 

\subsubsection{MMSE-Detection}
The Ziv-Zakai bound~\cite{Ziv_scalr_vector} establishes a connection between estimation accuracy and detection performance by offering a lower bound on the \ac{MSE} of parameter estimation. The bound is derived from the error probabilities associated with binary hypothesis tests between two possible parameter values~\cite{Ziv_scalr_vector,Ziv_vector_weight,Ziv_vector_hyperplane}.

\section{Score-Based Generative Models}\label{sec:ScoreBasis}
 
The metrics introduced in Sec.~\ref{sec:ISACmetrics} for evaluating the fundamental limitations of \ac{ISAC} systems highly rely on knowledge of the signal distributions. 
However, in practice, obtaining precise distributions is often infeasible.
While one could attempt to approximate these distributions using simplified models, 
these oversimplifications can result in a substantial mismatch between the assumed and true underlying distributions. As a result, these evaluations become unreliable as benchmarks for algorithm design and system optimization.

{To address this challenge, score-based generative models have emerged as a powerful solution.
Unlike traditional methods,  score-based generative models do not require explicit knowledge of the underlying distribution. Instead, they learn the score function, which is the gradient of the logarithm of the distribution, directly from available data samples. 
The learned score can then be used to generate new samples through Langevin dynamics~\cite{song2020score}. 
Beyond generative tasks, the learned score has recently been widely applied to solve inverse problems in estimation tasks~\cite{song2023pseudoinverse,daras2024survey}.
In this section, we introduce the fundamentals of learning the score function.}


\subsection{Score Function}

\begin{figure}
    \centering
    \includegraphics[width=0.8\linewidth]{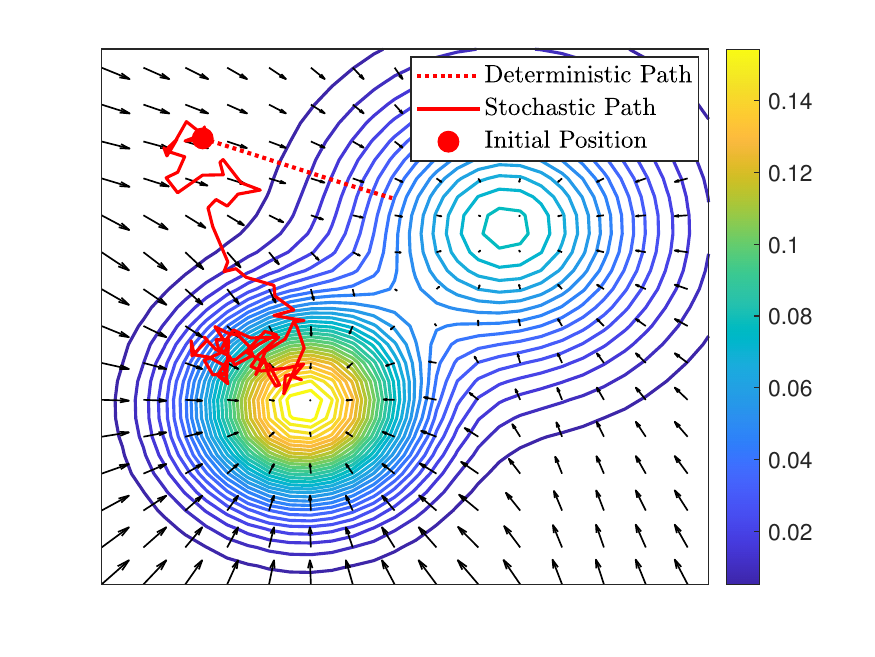}  
    \caption{Geometric interpretations of the score function.}
    \label{fig:gradient}
\end{figure}

The score function is defined as the gradient of the log-density, mathematically expressed as $\nabla_{\mathbf{x}}\log p(\mathbf{x})$, where $\mathbf{x} \in \mathbb{R}^D$ is a random vector and $p(\mathbf{x})$ is the \ac{PDF}.
Geometrically, the score function defines a vector field over the data space. 
Each point $\mathbf{x}$ in this high-dimensional space is associated with a vector $\nabla_{\mathbf{x}}\log p(\mathbf{x})$ pointing the steepest ascent direction in the log-density landscape.
This vector field can be visualized on contour maps of the density, as shown in Fig.~\ref{fig:gradient} for a Gaussian mixture model with two modes. The arrows represent the score vectors, showing how a data point moves along the contour lines toward the nearest mode or the high-density region.
This behavior allows for efficient exploration of the underlying structure of the data distribution.
One practical application that exploits this behavior is sampling via Langevin dynamics, where points iteratively move along the score vector field, combined with added noise to explore the space more broadly. Specifically, starting from any initial position, the points follow the score vectors toward the modes of the distribution, as depicted by the red trajectories in Fig.~\ref{fig:gradient}. The injected noise introduces stochasticity that prevents deterministic paths, thereby producing a more diverse and representative set of samples.


Building up on this, the score-based generative model learns the score function $\nabla_{\mathbf{x}}\log p(\mathbf{x})$ in order to sample from the distribution $p(\mathbf{x})$. This approach offers several advantages over other learning-based approaches that directly model the distribution $p(\mathbf{x})$~\cite{song2020score}. Due to the normalization constraint of $\int p(\mathbf{x}) \mathrm{d} \mathbf{x}=1$, directly learning $p(\mathbf{x})$ often imposes strong restrictions on the network architecture to ensure a tractable computation of the normalizing constant~\cite{hinton2002training}.
In contrast, the score function $\nabla_\mathbf{x} \log p({\mathbf{x}})$ is independent of the normalizing constant as the gradient of the constant with respect to $\mathbf{x}$ is zero. 
As a result, the score-based model learns an unconstrained function $\nabla_\mathbf{x} \log p({\mathbf{x}})$ that implicitly captures the structure of the data distribution without requiring explicit normalization. 
This makes the score-based model easier to train, particularly in high-dimensional settings~\cite{song2020score}.

\subsection{Score Matching}\label{sec:SM}
Given a dataset composed of independent and identically distributed (i.i.d.) samples $\mathcal{X}_{M}=\{\mathbf{x}_{m}\}_{m=1}^M$ drawn from an unknown distribution $p({\mathbf{x}})$, the score function can be approximated by a neural network  $\mathbf{s}_{\bm{\phi}}(\mathbf{x})$ parameterized by $\bm{\phi}$. 
The objective is to minimize the following \ac{ESM} loss~\cite{vincent2011connection}: 
\begin{align}\label{eq:loss}
    \mathcal{L}_{\rm ESM}(\bm{\phi})=\frac{1}{2}\mathbb{E}_{p(\mathbf{x})} \left[ \left\| \mathbf{s}_{\bm{\phi}}(\mathbf{x})-\nabla_\mathbf{x} \log p({\mathbf{x}})\right\| _2^2\right].
\end{align}
However, direct computation of the \ac{ESM} loss is impractical because the ground-truth score $\nabla_\mathbf{x} \log p({\mathbf{x}})$ is typically unknown. To address this, practical score matching techniques such as \ac{ISM}~\cite{hyvarinen2005estimation}, \ac{SSM}~\cite{song2020sliced}, and \ac{DSM}~\cite{vincent2011connection} have been developed, as detailed below.
 
\subsubsection{Implicit Score Matching}\label{sec:ISM}
\ac{ISM}~\cite{hyvarinen2005estimation} avoids the direct computation of the true score function by applying integration by parts to \eqref{eq:loss}. The ISM objective is given by
\begin{align}\label{eq:loss-ISM}
\begin{split}
\mathcal{L}_{\rm ISM}(\bm{\phi})&=\mathbb{E}_{p(\mathbf{x})} \left[ {\rm Tr}\left( \nabla_{\mathbf{x}}\,\mathbf{s}_{\bm{\phi}}(\mathbf{x}) \right) + \frac{1}{2}\left\| \mathbf{s}_{\bm{\phi}}(\mathbf{x}) \right\|_2^2 \right] ,
\end{split}
\end{align}
where $\nabla_{\mathbf{x}} \,\mathbf{s}_{\bm{\phi}}(\mathbf{x})$ denotes the Jacobian of $\mathbf{s}_{\bm{\phi}}(\mathbf{x})$.
Despite its conceptual simplicity, ISM is computationally demanding for high-dimensional data since calculating the trace term ${\rm Tr}\left( \nabla_{\mathbf{x}}\,\mathbf{s}_{\bm{\phi}}(\mathbf{x}) \right)$ involves multiple backpropagation passes through the neural network~\cite{song2020sliced}.

\subsubsection{Sliced Score Matching}
To mitigate the computational cost of the trace operator in \eqref{eq:loss-ISM}, \ac{SSM}~\cite{song2020sliced} provides a more scalable loss function based on Hutchinson's trace estimator~\cite{hutchinson1989stochastic}.
This estimator expresses the trace of a matrix as
\begin{align}\label{eq:trace}
    {\rm Tr}(\mathbf{A})=\mathbb{E}_{p(\mathbf{v})}\left[ \mathbf{v}^\mathsf{T} \mathbf{A} \mathbf{v}\right], \text{ for } \mathbf{A} \in \mathbb{R}^{D\times D},
\end{align}
where $p(\mathbf{v})$ is the \ac{PDF} of $\mathbf{v}$ that satisfies $\mathbb{E}[\mathbf{v}]=\mathbf{0}$ and $\mathbb{E}[\mathbf{v}\mathbf{v}^\mathsf{T}]=\mathbf{I}$. One possible choice of $p(\mathbf{v})$ is the
multivariate standard normal $\mathcal{N}(\mathbf{0},\mathbf{I})$. 
Applying \eqref{eq:trace} into \eqref{eq:loss-ISM}, the resulting \ac{SSM} loss is~\cite{song2020sliced} 
\begin{align}\label{eq:loss-SSM}
        \mathcal{L}_{\rm SSM}(\bm{\phi})
        &=
\mathbb{E}_{p(\mathbf{x})} \left[  \mathbb{E}_{p(\mathbf{v})}\left[\mathbf{v}^\mathsf{T}\nabla_{\mathbf{x}}\,\mathbf{s}_{\bm{\phi}}(\mathbf{x}) \mathbf{v} \right]+\frac{1}{2}  \left\| \mathbf{s}_{\bm{\phi}}(\mathbf{x})  \right\|_2^2\right]
        \nonumber\\& =
\mathbb{E}_{p(\mathbf{x})p(\mathbf{v})} \left[  \mathbf{v}^\mathsf{T}\nabla_{\mathbf{x}} (\mathbf{v}^\mathsf{T}\mathbf{s}_{\bm{\phi}}(\mathbf{x}))+\frac{1}{2}  \left\| \mathbf{s}_{\bm{\phi}}(\mathbf{x}) \right\| _2^2\right].
\end{align}
This function involves computing the gradient of a scalar function $\mathbf{v}^\mathsf{T}\mathbf{s}_{\bm{\phi}}(\mathbf{x})$ with respect to $\mathbf{x}$, which only requires a single backpropagation through the neural network.

\subsubsection{Denoising Score Matching}\label{subsec:DSM}
Instead of directly estimating the score of the original data distribution, \ac{DSM} introduces noise to the original data and estimates the score of the resulting noise-smoothed data distribution.
Denote the noise-smoothed data as $\mathbf{x}_t=\mathbf{x}+\sigma_t \mathbf{n}$, where $\mathbf{x}\in\mathcal{X}_{M}$ is the original clean data, $\sigma_t$ is the predefined noise level, and $\mathbf{n}\sim \mathcal{N}(\mathbf{0},\mathbf{I})$.
The corresponding smoothed data distribution is $p(\mathbf{x}_t)=\int p(\mathbf{x}_t|\mathbf{x}) p(\mathbf{x}) \mathrm{d} \mathbf{x}$, 
where $p(\mathbf{x}_t|\mathbf{x})=\mathcal{N}(\mathbf{x}_t;\mathbf{x},\sigma_t^2\mathbf{I})$ and
\begin{align}\label{eq:p(xt|x)}
    \nabla_{\mathbf{x}_t} \log p(\mathbf{x}_t|\mathbf{x})=-\frac{\mathbf{x}_t-\mathbf{x}}{\sigma_t^2}=-\frac{\mathbf{n}}{\sigma_t^2}.
\end{align}
To learn the {\em smoothed data score} $\nabla_{\mathbf{x}_t} \log p(\mathbf{x}_t)$, the ESM defines the loss function as
\begin{align}\label{eq:loss-ESM-noise}
\begin{split}
\mathcal{L}_{\rm ESM}(\bm{\phi},\sigma_t)&\!=\!
\frac{1}{2}\mathbb{E}_{p(\mathbf{x}_t)} \left[   \left\| \mathbf{s}_{\bm{\phi}}(\mathbf{x}_t,\sigma_t)-\!\nabla_{\mathbf{x}_t} \log p(\mathbf{x}_t) \right\|_2^2\right],
\end{split}
\end{align}
where $\mathbf{s}_{\bm{\phi}}(\mathbf{x}_t,\sigma_t)$ represents a neural network parameterized by $\bm \phi$, which takes both the noisy data $\mathbf{x}_t$ and the noise level $\sigma_t$ as inputs~\cite{song2019generative,song2020improved}.  Incorporating the noise level as an input enables the network to handle varying noise levels within a single unified model. 

The ESM loss in \eqref{eq:loss-ESM-noise} is not directly computable as we do not have access to the smoothed data distribution $p(\mathbf{x}_t)$. However, as shown in \cite{vincent2011connection}, $\mathcal{L}_{\rm ESM}(\bm{\phi},\sigma_t)$  is equivalent to the following \ac{DSM} loss:
\begin{align}\label{eq:loss-DSM}
\mathcal{L}_{\rm DSM}(\bm{\phi},\sigma_t)&=
\frac{1}{2}\mathbb{E}_{p(\mathbf{x}_t,\mathbf{x})} \left[ \left\| \mathbf{s}_{\bm{\phi}}(\mathbf{x}_t,\sigma_t)-\nabla_{\mathbf{x}_t} \log p(\mathbf{x}_t|{\mathbf{x}}) \right\| _2^2\right]
\nonumber\\&\overset{(a)}{=}\frac{1}{2}\mathbb{E}_{p(\mathbf{x})p(\mathbf{n})} \left[\left\| \mathbf{s}_{\bm{\phi}}(\mathbf{x}_t,\sigma_t)+\frac{\mathbf{n}}{\sigma_t^2}  \right\|_2^2\right],
\end{align}
where (a) is from \eqref{eq:p(xt|x)}.
This formulation interprets
$\mathbf{s}_{\bm{\phi}}(\mathbf{x}_t,\sigma_t)$ as a noise predictor or denoiser, which predicts the scaled noise $\frac{\mathbf{n}}{\sigma_t^2}$ from the noisy data $\mathbf{x}_t=\mathbf{x}+\sigma_t \mathbf{n}$, effectively recovering the clean data $\mathbf{x}$.
Therefore, by injecting noise into the original data, DSM thus provides a computable and practical loss function for learning the smoothed data score $\nabla_{\mathbf{x}_t} \log p(\mathbf{x}_t)$.
In addition to its computational practicality, DSM overcomes the difficulty of accurate score estimation in low data density regions, where few data samples are available for computing the score matching objective~\cite{song2019generative,song2020improved}. This is because noise injection can populate low data density regions with noisy samples, thereby improving the accuracy of the estimated score in these regions.

The above training procedure assumes a single noise level $\sigma_t$.
Generalizing it to multiple noise levels $\{\sigma_t\}_{t=1}^{T}$, the corresponding loss is~\cite{song2019generative}
\begin{align}\label{eq:loss-NCSN}
    \mathcal{L}_{{\rm DSM},T}(\bm{\phi})=\frac{1}{T}\sum_{t=1}^{T} \lambda(\sigma_t) \mathcal{L}_{\rm DSM}(\bm{\phi},\sigma_t),
\end{align}
where $\sigma_t$ $(t=1,...,T)$ denotes the $t$-th noise level, and the weight $\lambda(\sigma_t)$  is empirically set as $\sigma_t^2$. 
 
\subsection{Conditional Score}\label{sec:CondScore}

In the previous subsection, we focused on modeling the unconditional/prior distribution $p(\mathbf{x})$.
Modeling the conditional/posterior distribution $p(\mathbf{x}|\mathbf{y})$ is also crucial in many practical scenarios as it reflects how the variable $\mathbf{x}$ is influenced or constrained by the observation $\mathbf{y}$.
For example, the \ac{MMSE} estimator in \eqref{eq:mmse_mean} is determined by the posterior mean, and the computation of Bayesian FIM in \eqref{eq:BFIM-posterior} requires access to the posterior distribution.

Consider a general form of the measurement model as
\begin{align}
\label{eq:datamodel}
    \mathbf{y}=\mathcal{A}(\mathbf{x})+\mathbf{z},
\end{align}
where $\mathcal{A}$ is the measurement operator and $\mathbf{z}$ is the measurement noise.  
Applying Bayes' rule, the {\em conditional/posterior score} $\nabla_{\mathbf{x}}\log p(\mathbf{x}|\mathbf{y})$ 
can be decomposed as
\begin{align}\label{eq:cond-score}
    \underbrace{\nabla_{\mathbf{x}}\log p(\mathbf{x}|\mathbf{y})}_{\substack{\rm conditional~score \\\rm (posterior~score)}}= 
    \underbrace{\nabla_{\mathbf{x}}\log p(\mathbf{x})}_{\substack{\rm unconditional~score\\\rm (prior~score)}} + \underbrace{\nabla_{\mathbf{x}}\log p(\mathbf{y}|\mathbf{x}),}_{\substack{\rm measurement~score\\\rm (Fisher~ score)}}
\end{align}
where $\nabla_{\mathbf{x}}\log p(\mathbf{x})$ is termed as {\em unconditional/prior score} related to the prior distribution and $\nabla_{\mathbf{x}}\log p(\mathbf{y}|\mathbf{x})$ is termed as {\em measurement/Fisher score} related to the measurement distribution (or likelihood function). 
In the following, we introduce \ac{ISM} and \ac{DSM} techniques for learning the conditional/posterior score. 
Since the adaptation from \ac{ISM} to \ac{SSM} can be achieved based on Hutchinson's trace estimator in \eqref{eq:trace}, we omit the details here for brevity. 
Additionally, for each score matching technique, we present two approaches for learning the posterior score: (i) {\em posterior approach}, which directly learns the posterior score, and (ii) {\em measurement-prior approach}, which estimates the posterior score by combining a pre-trained prior score with information from the measurement model.

\subsubsection{ISM for Conditional Score}\label{sec:cond-ISM}
The {\em posterior approach} directly learns the conditional score by extending the unconditional ISM formulation in Sec.~\ref{sec:ISM}.  
Specifically, consider a neural network $\mathbf{s}_{\bm \phi}(\mathbf{x}|\mathbf{y})$ parameterized by $\bm \phi$, which takes both the data $\mathbf{x}$ and the condition $\mathbf{y}$ as inputs. Similar to \eqref{eq:loss-ISM}, the ISM defines the loss for learning the conditional score $\nabla_{\mathbf{x}}\log p(\mathbf{x}|\mathbf{y})$ as~\cite{habi2025learned} 
\begin{align}\label{eq:loss-cond-ISM}
\begin{split}
\mathcal{L}_{\rm ISM}^{\rm cond}(\bm{\phi})&=\mathbb{E}_{p(\mathbf{x},\mathbf{y})} \left[ {\rm Tr}\left( \nabla_{\mathbf{x}}\,\mathbf{s}_{\bm{\phi}}(\mathbf{x}|\mathbf{y}) \right) + \frac{1}{2}\left\| \mathbf{s}_{\bm{\phi}}(\mathbf{x}|\mathbf{y}) \right\|_2^2 \right].
\end{split}
\end{align}
Moreover, when both the unconditional and conditional scores are needed, training separate neural networks can be inefficient. A common practice is to use a single network ${\mathbf{s}}_{\bm \phi}(\mathbf{x}|\mathbf{y})$ that outputs the unconditional score  $\nabla_{\mathbf{x}}\log p(\mathbf{x})$ by setting the condition input as empty ($\mathbf{y}=\emptyset$), and the conditional score $\nabla_{\mathbf{x}}\log p(\mathbf{x}|\mathbf{y})$ by using non-empty condition input.
During the training process of ${\mathbf{s}}_{\bm \phi}(\mathbf{x}|\mathbf{y})$, the condition $\mathbf{y}$ is randomly set as $\emptyset$ with probability $p_{\rm uncond}$ (set as hyperparameter), allowing the network to jointly learn the unconditional and conditional scores.

An alternative is the {\em measurement-prior} approach, which computes the conditional score using a pre-trained unconditional score and the measurement model information. Suppose an unconditional score has been learned using \ac{ISM}, denoted by 
\begin{align}\label{eq:s-ISM}
    \mathbf{s}_{\bm{\phi}_{\rm ISM}}(\mathbf{x}) \approx \nabla_{\mathbf{x}} \log p(\mathbf{x}).
\end{align}
If the {\em measurement distribution $p(\mathbf{y}|\mathbf{x})$ is known}, the conditional score can be readily estimated based on \eqref{eq:s-ISM} and \eqref{eq:cond-score}~\cite{crafts2024bayesian}. 
For example, when the measurement noise in \eqref{eq:datamodel} follows $\mathcal{N}(\mathbf{z};\mathbf{0},\sigma_z^2\mathbf{I})$, the measurement score simplifies to
\begin{align}\label{eq:p(y|x)_gaussian}
    \nabla_{\mathbf{x}}\log p(\mathbf{y}|\mathbf{x}) =-\frac{1}{\sigma_z^2} \nabla_{\mathbf{x}} \left\| \mathbf{y}-\mathcal{A}(\mathbf{x})\right\|^2_2.
\end{align}
Substituting into \eqref{eq:p(y|x)_gaussian} and \eqref{eq:s-ISM} into \eqref{eq:cond-score} provides a direct way to estimate  $\nabla_{\mathbf{x}}\log p(\mathbf{x}|\mathbf{y})$.

In more complicated cases, we only know the measurement operator $\mathcal{A}(\cdot)$ while the likelihood $p(\mathbf{y}|\mathbf{x})$ is not explicitly available. 
Based on $p(\mathbf{y}|\mathbf{x})=p(\mathbf{y}|\mathcal{A}(\mathbf{x}))$ and the chain rule, the measurement score can be expressed as~\cite{habi2025learned} 
\begin{align}\label{eq:PeNN}
    \nabla_{\mathbf{x}}\log p(\mathbf{y}|\mathbf{x})=\left(\frac{\partial \mathcal{A}(\mathbf{x})}{\partial \mathbf{x}}\right)^\mathsf{T} \nabla_{\bm \tau}\log p(\mathbf{y}|\bm \tau)\big|_{\bm \tau=\mathcal{A}(\mathbf{x})},
\end{align}
where $\left(\frac{\partial \mathcal{A}(\mathbf{x})}{\partial \mathbf{x}}\right)^\mathsf{T}$ is known and a neural network model $\mathbf{s}_{\phi}(\mathbf{y}|\bm\tau)$ can be employed to learn $\nabla_{\bm \tau}\log p(\mathbf{y}|\bm \tau)$. Then, $\nabla_{\mathbf{x}}\log p(\mathbf{y}|\mathbf{x})$ can be modeled as~\cite{habi2025learned} 
 \begin{align}
 \mathbf{s}_{\phi}(\mathbf{y}|\mathbf{x})= \left(\frac{\partial \mathcal{A}(\mathbf{x})}{\partial \mathbf{x}}\right)^\mathsf{T} \mathbf{s}_{\phi}(\mathbf{y}|\bm\tau)\big|_{\bm \tau=\mathcal{A}(\mathbf{x})},
\end{align}
Note that $\nabla_{\mathbf{x}}\log p(\mathbf{y}|\mathbf{x})$ differs from a standard score function since it is the gradient of the log-likelihood with respect to $\mathbf{x}$ instead of $\mathbf{y}$.
A new score matching technique, \ac{FSM}~\cite{habi2025learned}, is used for training $\mathbf{s}_{\phi}(\mathbf{y}|\mathbf{x})$.
\ac{FSM} applies integration by parts to transform the intractable ESM loss  
\begin{align}
    \frac{1}{2}\mathbb{E}_{p(\mathbf{x},\mathbf{y})} \left[\left\|  \mathbf{s}_{\bm{\phi}}(\mathbf{y}|\mathbf{x})-\nabla_\mathbf{x} \log p(\mathbf{y}|{\mathbf{x}}) \right\|_2^2\right]
\end{align}  
into a tractable objective. The resulting FSM loss is given by~\cite{habi2025learned}
\begin{align}\label{eq:loss-FSM}
\mathcal{L}_{\rm FSM}(\bm{\phi})&=\mathbb{E}_{p(\mathbf{y})p(\mathbf{x})} \bigg[ {\rm Tr}\left( \nabla_{\mathbf{x}} \mathbf{s}_{\bm{\phi}}(\mathbf{y}|\mathbf{x}) \right) + \frac{1}{2}  \left\| \mathbf{s}_{\bm{\phi}}(\mathbf{y}|\mathbf{x}) \right\|_2^2 +\nonumber \\&\qquad\mathbf{s}_{\bm{\phi}}(\mathbf{y}|\mathbf{x})^\mathsf{T} \nabla_{\mathbf{x}} \log p(\mathbf{x}) \bigg],
\end{align} 
Since the true prior score $\nabla_{\mathbf{x}} \log p(\mathbf{x})$ is generally unknown, the pre-trained unconditional score $\mathbf{s}_{\bm{\phi}_{\rm ISM}}(\mathbf{x})$ in \eqref{eq:s-ISM} 
can be used as a surrogate.    
This substitution makes the objective in~\eqref{eq:loss-FSM} feasible for practical training.
Denote the learned measurement score by $\mathbf{s}_{\phi_{\rm FSM}}(\mathbf{y}|\mathbf{x})$.
Based on \eqref{eq:cond-score}, the conditional score can then be approximated as $\mathbf{s}_{\bm{\phi}_{\rm ISM}}(\mathbf{x}) + \mathbf{s}_{\phi_{\rm FSM}}(\mathbf{y}|\mathbf{x})$.

\subsubsection{DSM for Conditional Score}\label{sec:cond-DSM}
 
The \ac{DSM}-based {\em posterior}  approach directly learns the smoothed conditional score $ \nabla_{\mathbf{x}_t} \log p(\mathbf{x}_t|\mathbf{y})$, where $\mathbf{x}_t=\mathbf{x}+\sigma_t  \mathbf{n}$ is the noise-smoothed data, $\mathbf{x}\in\mathcal{X}_M$ is the original clean data, and $\mathbf{y}$ is given in \eqref{eq:datamodel}. 
Following the DSM technique introduced in Sec.~\ref{subsec:DSM},
the DSM loss for learning $ \nabla_{\mathbf{x}_t} \log p(\mathbf{x}_t|\mathbf{y})$ is
\begin{align}\label{eq:loss-DSM-posterior}
\begin{split}
&\mathcal{L}_{\rm DSM}^{{\rm cond}}(\bm{\phi},\sigma_t)
\!=\!\!\frac{1}{2}\mathbb{E}_{p(\mathbf{x},\mathbf{y})p(\mathbf{n})} \!\!\left[\left\| {\mathbf{s}}_{\bm{\phi}}(\mathbf{x}_t,\sigma_t|\mathbf{y}) \!+\!\frac{\mathbf{n}}{\sigma_t^2}  \right\|_2^2\right]\!,
\end{split}
\end{align}
where ${\mathbf{s}}_{\bm \phi}(\mathbf{x}_t,\sigma_t|\mathbf{y})$ represents a neural network with the inputs including $\mathbf{x}_t$, $\sigma_t$, and $\mathbf{y}$.
Moreover, similar to the posterior approach in the \ac{ISM}-based conditional score, the single neural network ${\mathbf{s}}_{\bm \phi}(\mathbf{x}_t,\sigma_t|\mathbf{y})$ can jointly learn both the smoothed unconditional score $\nabla_{\mathbf{x}_t} \log p(\mathbf{x}_t)$ and the smoothed 
conditional score $ \nabla_{\mathbf{x}_t} \log p(\mathbf{x}_t|\mathbf{y})$. This unified modeling method is known as {\em classifier-free guidance}~\cite{ho2022classifier}.

In the \ac{DSM}-based {\em measurement-prior approach}, instead of learning the unconditional score from scratch, we use a pre-trained model to approximate the smoothed unconditional score, i.e., 
\begin{align}\label{eq:score-NCSN}
    \mathbf{s}_{\bm{\phi}_{\rm DSM}}(\mathbf{x}_t,\sigma_t)  \approx \nabla_{\mathbf{x}_t} \log p(\mathbf{x}_t).
\end{align}
Using Bayes' rule, the smoothed conditional score can be decomposed as
\begin{align}\label{eq:conid-score-dsm}
    \nabla_{\mathbf{x}_t}\log p(\mathbf{x}_t|\mathbf{y})= \nabla_{\mathbf{x}_t}\log p(\mathbf{x}_t) + \nabla_{\mathbf{x}_t}\log p(\mathbf{y}|\mathbf{x}_t).
\end{align}
However, different from \eqref{eq:cond-score}, the second term on the right-hand side of \eqref{eq:conid-score-dsm} is difficult to acquire even when the likelihood $p(\mathbf{y}|\mathbf{x})$ is known. 
Specifically, $p(\mathbf{y}|\mathbf{x}_t)$ is given by~\cite{chung2022diffusion}  
\begin{align}\label{eq:p(y|xt)}
\begin{split}
    p(\mathbf{y}|\mathbf{x}_t)&=\int p(\mathbf{y}|\mathbf{x},\mathbf{x}_t) p(\mathbf{x}|\mathbf{x}_t) \mathrm{d} \mathbf{x}
    \\&\overset{(a)}{=}\int p(\mathbf{y}|\mathbf{x}) p(\mathbf{x}|\mathbf{x}_t) \mathrm{d} \mathbf{x},    
\end{split}
\end{align}
where (a) holds because $\mathbf{x}_t$ and $\mathbf{y}$ are conditionally independent on $\mathbf{x}$. However, $p(\mathbf{x}|\mathbf{x}_t)$ is generally intractable, leading to unavailable $p(\mathbf{y}|\mathbf{x}_t)$.

To approximate the second term $\nabla_{\mathbf{x}_t}\log p(\mathbf{y}|\mathbf{x}_t)$ in \eqref{eq:conid-score-dsm}, a line of work is to train a classifier neural network $p_{\bm \psi}(\mathbf{y}|\mathbf{x}_t)$ parametrized by $\bm \psi$~\cite{dhariwal2021diffusion}. Taking the gradient of the learned classifier yields $\nabla_{\mathbf{x}_t} \log p_{\bm \psi}(\mathbf{y}|\mathbf{x}_t) $, which serves as an approximation for $ \nabla_{\mathbf{x}_t} \log p(\mathbf{y}|\mathbf{x}_t)$. 
With the pre-trained smoothed unconditional score and the classifier, the conditional score can then be approximated based on \eqref{eq:conid-score-dsm}.
This method is known as {\em classifier guidance}~\cite{dhariwal2021diffusion}. 
However, a drawback of classifier guidance is its reliance on a separately learned classifier~\cite{daras2024survey}.

Another line of work~\cite{chung2022diffusion,song2023pseudoinverse} provides explicit approximations for $\nabla_{\mathbf{x}_t}\log p(\mathbf{y}|\mathbf{x}_t)$ based on \eqref{eq:p(y|xt)},  
where the likelihood $p(\mathbf{y}|\mathbf{x})$ is required to be known, and $\hat{p}(\mathbf{x}|\mathbf{x}_t) $ serves as an approximation of  $p(\mathbf{x}|\mathbf{x}_t)$. 
Different forms of $\hat{p}(\mathbf{x}|\mathbf{x}_t) $ have been proposed in the literature. 
For example, $\hat{p}(\mathbf{x}|\mathbf{x}_t)$ is modeled as a delta function $\delta(\mathbf{x}-\hat{\mathbf{x}}_{0|t})$~\cite{chung2022diffusion} or a Gaussian distribution $\mathcal{N}(\mathbf{x};\hat{\mathbf{x}}_{0|t},\zeta_t \mathbf{I})$ with a hyperparameter $\zeta_t$~\cite{song2023pseudoinverse}. Here, $\hat{\mathbf{x}}_{0|t}$ is
an approximation of $\mathbb{E}_{p(\mathbf{x}|\mathbf{x_t})}[\mathbf{x}]$, which can be computed from $\mathbf{s}_{\bm{\phi}_{\rm DSM}}(\mathbf{x}_t,\sigma_t)$. Specifically, Tweedie's formula~\cite{robbins1992empirical} gives 
\begin{align}\label{eq:Tweedie}
\mathbb{E}_{p(\mathbf{x}|\mathbf{x}_t)}[\mathbf{x}]
=\mathbf{x}_t+\sigma_t^2\nabla_{\mathbf{x}_t} \log p(\mathbf{x}_t),
\end{align}
which leads to the approximation
\begin{align}\label{eq:x{0|t}}
\hat{\mathbf{x}}_{0|t}=\mathbf{x}_t+\sigma_t^2 \mathbf{s}_{\bm{\phi}_{\rm DSM}}(\mathbf{x}_t,\sigma_t).
\end{align}

Explicit approximations for $\nabla_{\mathbf{x}_t}\log p(\mathbf{x}_t|\mathbf{y})$ can also be derived using the conditional version of Tweedie's formula~\cite{robbins1992empirical}:
\begin{align}\label{eq:conid-score-opt}
    \nabla_{\mathbf{x}_t}\log p(\mathbf{x}_t|\mathbf{y})=\frac{\mathbb{E}_{p(\mathbf{x}|\mathbf{x}_t,\mathbf{y})}[\mathbf{x}]-\mathbf{x}_t}{\sigma_t^2}.
\end{align}
The key to evaluating $\nabla_{\mathbf{x}_t}\log p(\mathbf{x}_t|\mathbf{y})$ in \eqref{eq:conid-score-opt} lies in estimating $\mathbb{E}_{p(\mathbf{x}|\mathbf{x}_t,\mathbf{y})}[\mathbf{x}]$.
The approximation of  $\mathbb{E}_{p(\mathbf{x}|\mathbf{x}_t,\mathbf{y})}[\mathbf{x}]$, denoted by $\hat{\mathbf{x}}_{0|t,\mathbf{y}}$, can be obtained through various optimization objectives. 
For example, when dealing with a linear measurement operator $\mathcal{A}(\mathbf{x})=\mathbf{A}\mathbf{x}$, $\hat{\mathbf{x}}_{0|t,\mathbf{y}}$ can be computed using the hard data-consistency step in \eqref{eq:hard}~\cite{wang2022zero} or the soft data-consistency step in \eqref{eq:soft}~\cite{chung2023decomposed,zhu2023denoising}, i.e.,
\begin{subequations}
\begin{align}\label{eq:hard}
\hat{\mathbf{x}}_{0|t,\mathbf{y}}&=\arg\min  \left\| \mathbf{x}-\hat{\mathbf{x}}_{0|t} \right\|_2^2
\nonumber\\& \quad\qquad \text{\rm s.t.  } \mathbf{x}\in \arg\min  \left\| \mathbf{y}-\mathbf{A}\mathbf{x} \right\|_2^2,
\\\label{eq:soft}
\hat{\mathbf{x}}_{0|t,\mathbf{y}}&=\arg\min \gamma_t  \left\| \mathbf{x}-\hat{\mathbf{x}}_{0|t} \right\|_2^2+  \left\|\mathbf{y}-\mathbf{A}\mathbf{x} \right\| _2^2,
\end{align}     
\end{subequations}
where $\gamma_t>0$ is a hyperparameter and $\hat{\mathbf{x}}_{0|t}$ is obtained based on the pre-trained prior score, as given in \eqref{eq:x{0|t}}.

\section{Score for \ac{ISAC} Performance Evaluation}\label{sec:Score-ISAC}

This section utilizes the score-based generative models to characterize the fundamental metrics in \ac{ISAC} systems, including the \ac{FIM}, \ac{MMSE}, \ac{MI}, and \ac{KLD}.
The proposed scoring \ac{ISAC} framework, as shown in Fig.~\ref{fig:Framework}, 
allows for a more practical assessment of \ac{ISAC} systems in a data-driven approach.

\subsection{Scoring FIM}\label{sec:scoreFIM}
Consider the general parameter estimation model given in \eqref{eq:estimation}, i.e., $\mathbf{y}={\bm f}(\bm{\theta})+\mathbf{z}$. To evaluate Bayesian \ac{FIM}, we employ ISM-based posterior and measurement-prior approaches.
Note that we replace $\mathbf{x}$ with $\bm \theta$ in score-based networks and loss functions defined in Sec.~\ref{sec:ScoreBasis}.  
Moreover, we denote $\{\bm \theta_m\}_{m=1}^M$ as i.i.d. samples from $p(\bm \theta)$, and $\{\mathbf{y}_{m,\ell}\}_{\ell=1}^L$ as i.i.d. samples from $p(\mathbf{y}|\bm\theta_m)$. 

\subsubsection{Posterior Approach With ISM} In this approach, we utilize the posterior-form Bayesian FIM in \eqref{eq:BFIM-posterior}, which requires the conditional score $\nabla_{\bm \theta}\log p(\bm \theta|\mathbf{y})$.  
Denote $\mathbf{s}_{\phi^{\rm cond}_{\rm ISM}}(\bm\theta|\mathbf{y})$ as the learned conditional score by minimizing the loss $\mathcal{L}_{\rm ISM}^{{\rm cond}}(\bm{\phi})$ in \eqref{eq:loss-cond-ISM}. Then, the Bayesian FIM $\mathbf{J}_{\rm b}$ can be estimated by~\cite{habi2025learned}
\begin{align}\label{eq:BFM-P-ISM}
        \tilde{\mathbf{J}}_{\rm b}
        &= \frac{1}{ML}\sum_{m=1}^{M} \sum_{\ell=1}^{L} [\mathbf{s}_{\phi^{\rm cond}_{\rm ISM}}(\bm\theta_m|\mathbf{y}_{m,\ell}) \mathbf{s}_{\phi^{\rm cond}_{\rm ISM}}(\bm\theta_m|\mathbf{y}_{m,\ell})^\mathsf{T}].
    \end{align}
where the sample mean is used to approximate the expectation $\mathbb{E}_{p(\mathbf{y},\bm{\theta})}[\cdot]$.

\subsubsection{Measurement-Prior Approach With ISM}\label{subsec:ISM-meas-prior}
In this approach, we utilize the decomposition in \eqref{eq:BFIM-Bayes}, which expresses the Bayesian FIM as the sum of the prior FIM $\mathbf{J}_{\rm p}$ and the data FIM $\mathbf{J}_{\rm d}$. This leads to the approximation:  
\begin{align}
    \hat{\mathbf{J}}_{\rm b}=\hat{\mathbf{J}}_{\rm p}+\hat{\mathbf{J}}_{\rm d},
\end{align}
where $\hat{\mathbf{J}}_{\rm p}$ and $\hat{\mathbf{J}}_{\rm d}$ are estimators of $\mathbf{J}_{\rm p}$ and $\mathbf{J}_{\rm d}$, respectively, as detailed below.

Using the learned unconditional score via ISM, $\mathbf{s}_{\bm \phi_{\rm ISM}}(\bm\theta)$, to approximate $ \nabla_{\bm \theta}\log p(\bm \theta)$, the prior FIM can be estimated by
\begin{align}\label{eq:PFIM-ISM}
\begin{split}
    \hat{\mathbf{J}}_{\rm p} 
        = \frac{1}{M}\sum_{m=1}^{M}[\mathbf{s}_{\bm \phi_{\rm ISM}}(\bm\theta_m) \mathbf{s}_{\bm \phi_{\rm ISM}}(\bm\theta_m)^\mathsf{T}],
\end{split}
    \end{align}
where the sample mean is used to approximate the expectation $\mathbb{E}_{p(\bm{\theta})}[\cdot]$. 

The evaluation of the data FIM requires the Fisher score $\nabla_{\bm \theta}\log p(\mathbf{y}|\bm \theta)$. 
We consider the following two cases where the likelihood $p(\mathbf{y}|\bm \theta)$ is {\em known} and {\em unknown}.
First, when $p(\mathbf{y}|\bm \theta)$ is {\em known}, e.g., the Gaussian distribution in \eqref{eq:p(y|x)_gaussian}, the measurement score $\nabla_{\bm \theta}\log p(\mathbf{y}|\bm \theta)$ is available. Then, $\mathbf{J}_{\rm d}$ can be evaluated by~\cite{crafts2024bayesian}
\begin{align}\label{eq:BFIM-MP-ISM}
\begin{split}
\hat{\mathbf{J}}_{\rm d} &  
 = \frac{1}{ML}\sum_{m=1}^{M} \sum_{\ell=1}^{L} [\nabla_{\bm \theta_m}\log p(\mathbf{y}_{m,\ell}|\bm \theta_m) \nabla_{\bm \theta_m}\log p(\mathbf{y}_{m,\ell}|\bm \theta_m)^\mathsf{T}],  
\end{split}
\end{align}
where the sample mean is used to approximate the expectation $\mathbb{E}_{p(\mathbf{y},\bm{\theta})}[\cdot]$.
Second, when $p(\mathbf{y}|\bm \theta)$ is {\em unknown}, e.g., {due to the presence of both interference and noise}, $\nabla_{\bm \theta}\log p(\mathbf{y}|\bm \theta)$ is unavailable. 
Let $\mathbf{s}_{\bm \phi_{\rm FSM}}(\mathbf{y}|\bm \theta)$ be the learned measurement score via FSM in \eqref{eq:loss-FSM}. Then, $\mathbf{J}_{\rm d}$ can be estimated by~\cite{habi2025learned}
\begin{align}\label{eq:DFIM-FSM}
\begin{split}
        \hat{\mathbf{J}}_{\rm d}&
        =\frac{1}{ML}\sum_{m=1}^{M} \sum_{\ell=1}^{L} [\mathbf{s}_{\bm \phi_{\rm FSM}}(\mathbf{y}_{m,\ell}|\bm\theta_m) \mathbf{s}_{\bm \phi_{\rm FSM}}(\mathbf{y}_{m,\ell}|\bm\theta_m)^\mathsf{T}].
\end{split}
\end{align}

\subsection{Scoring MMSE}\label{sec:scoreMMSE}
Let $\mathbf{x}$ and $\mathbf{y}$ be two random variables.
The MMSE estimator of $\mathbf{x}$ given $\mathbf{y}$ is the posterior mean $\mathbb{E}\left[\mathbf{x}|\mathbf{y}\right]$.
The corresponding MMSE is expressed as
\begin{align}
    \mathbb{E}\left[ \left\| \mathbf{x}-\mathbb{E}\left[\mathbf{x}|\mathbf{y}\right] \right\|^2_2\right].
\end{align}
However, computing the MMSE estimator, or evaluating the MMSE, is often intractable in practice, particularly when the joint distribution $p(\mathbf{x}, \mathbf{y})$ is high-dimensional, complex, or only implicitly known.
To address this challenge, recent advances in score-based generative modeling offer a promising alternative to calculate the MMSE estimator without performing explicit integration \cite{cai2025score, cai2025massive_connectivity, xue2024score, wang2025traversing, wentao2024score}.
With the score-based MMSE estimator, the associated MMSE can be evaluated accordingly.
We discuss the score-based MMSE estimators for different measurement models in the following.

\subsubsection{AWGN Observation Model}
Consider the measurement model $\mathbf{y} = \mathbf{x} + \sigma \mathbf{n}$, where $\mathbf{n}\sim \mathcal{N}(\mathbf{0}, \mathbf{I})$.
The MMSE estimator admits a closed-form expression known as Tweedie's formula:
\begin{align}
\mathbb{E}\left[\mathbf{x}|\mathbf{y}\right]=\mathbf{y} + \sigma^2\nabla_{\mathbf{y}} \log p(\mathbf{y}).
\end{align}

\subsubsection{Linear Inverse Problem}
For the measurement model $\mathbf{y} = \mathbf{A}\mathbf{x} + \sigma \mathbf{n}$, Tweedie's formula is no longer directly applicable, as the linear mixing introduced by $\mathbf{A}$ complicates the posterior distribution $p(\mathbf{x}|\mathbf{y})$. Consequently, computing the MMSE estimator becomes significantly more challenging.
Building upon the turbo compressive sensing (Turbo-CS) framework \cite{ma2015turbo_cs}, which is derived from the expectation propagation (EP) principle \cite{EP2001Minka}, a score-based turbo message passing (STMP) algorithm was proposed in \cite{cai2025score} for posterior mean calculation.
Specifically, ref. \cite{cai2025score} proposed the integration of score-based generative models into message passing algorithms as a plug-and-play denoiser.
STMP iterates between a linear estimator and a score-based MMSE denoiser. 

The linear estimator produces a linear estimate of $\mathbf{x}$ by combining the likelihood $p(\mathbf{y}|\mathbf{x})$ with a pseudo Gaussian prior of the form $p(\mathbf{x}) = \mathcal{N}(\mathbf{x}; \mathbf{x}_{\sf A}^{\rm pri}, v_{\sf A}^{\rm pri}\mathbf{I})$. 
As a result, the posterior mean and variance of the linear estimator are respectively given by
\begin{align}
	\mathbf{x}_{\sf A}^{\rm post} &= \mathbf{x}_{\sf A}^{\rm pri} + v_{\sf A}^{\rm pri}\mathbf{A}^{\sf T} \big(v_{\sf A}^{\rm pri}\mathbf{A}\mathbf{A}^{\sf T} + \sigma^2\mathbf{I}\big)^{-1} \big(\mathbf{y} - \mathbf{A} \mathbf{x}_{\sf A}^{\rm pri}\big), \label{lmmse_mean}\\
	v_{\sf A}^{\rm post} &= v_{\sf A}^{\rm pri} - \frac{(v_{\sf A}^{\rm pri})^2}{N} \mathrm{tr} \left(\mathbf{A}^{\sf T} \big(v_{\sf A}^{\rm pri}\mathbf{A} \mathbf{A}^{\sf T} + \sigma^2 \mathbf{I}\big)^{-1}\mathbf{A}\right), \label{lmmse_variance}
\end{align}
which is essentially the LMMSE estimate of $\mathbf{x}$.
The extrinsic variance and mean are respectively given by
\begin{align}
    v_{\sf A}^{\rm ext} &= \left(\frac{1}{v_{\sf A}^{\rm post}} - \frac{1}{v_{\sf A}^{\rm pri}}\right)^{-1}, \\
    \mathbf{x}_{\sf A}^{\rm ext} &= v_{\sf A}^{\rm ext} \left(\frac{\mathbf{x}_{\sf A}^{\rm post}}{v_{\sf A}^{\rm post}} - \frac{\mathbf{x}_{\sf A}^{\rm pri}}{v_{\sf A}^{\rm pri}} \right).
\end{align}
The extrinsic output of the linear estimator serves as the input to the score-based MMSE denoiser: $\mathbf{x}_{\sf B}^{\rm pri} = \mathbf{x}_{\sf A}^{\rm ext}$, $v_{\sf B}^{\rm pri} = v_{\sf A}^{\rm ext}$.

The denoiser produces a nonlinear estimate of $\mathbf{x}$ exploiting the prior $p(\mathbf{x})$ and a pseudo likelihood $p(\mathbf{x}|\mathbf{x}_{\sf B}^{\rm pri}) = \mathcal{N}(\mathbf{x}; \mathbf{x}_{\sf B}^{\rm pri}, v_{\sf B}^{\rm pri}\mathbf{I})$.
In other words, the input to the denoiser is assumed to be an AWGN observation of the ground truth, i.e., $\mathbf{x}_{\sf B}^{\rm pri} = \mathbf{x} + \mathbf{w}$, where $\mathbf{w}\sim \mathcal{N}(\mathbf{0}, v_{\sf B}^{\rm pri}\mathbf{I})$.
This assumption can be justified by the state evolution (SE) analysis in \cite{se_analysis2018nips}, showing that the error vector $\mathbf{x}_{\sf B}^{\rm pri}  - \mathbf{x}$ is i.i.d. Gaussian in the large system limit.
Based on the equivalent AWGN model, the posterior mean of the MMSE denoiser is expressed as
\begin{align}
    \mathbf{x}_{\sf B}^{\rm post} = \mathbb{E}\big[\mathbf{x}_{\sf B}|\mathbf{x}_{\sf B}^{\rm pri}\big]
    = \mathbf{x}_{\sf B}^{\rm pri} + v_{\sf B}^{\rm pri} \nabla_{\mathbf{x}_{\sf B}^{\rm pri}} \log p\left(\mathbf{x}_{\sf B}^{\rm pri}\right).
\end{align}
The posterior variance of the MMSE denoiser can be obtained in different ways.
If the second-order score function is available, the posterior variance can be calculated through Tweedie's covariance formula \cite{efron2011tweedie} by \cite{cai2025score, cai2025massive_connectivity}
\begin{align}
    v_{\sf B}^{\rm post} = v_{\sf B}^{\rm pri} + \frac{(v_{\sf B}^{\rm pri})^2}{N} \mathrm{tr} \left( \nabla_{\mathbf{x}_{\sf B}^{\rm pri}}^2 \log p\left(\mathbf{x}_{\sf B}^{\rm pri}\right) \right).
\end{align}
Otherwise, one can approximate the posterior variance by \cite{philip2013EM_GM}
\begin{align}
    v_{\sf B}^{\rm post} =\frac{\|\mathbf{y} - \mathbf{A}\mathbf{x}_{\sf B}^{\rm post}\|_2^2 - M \sigma^2}{\left\|\mathbf{A}\right\|_F^2},
\end{align}
under the fact that the estimation error $\mathbf{w}$ is uncorrelated with the noise $\mathbf{n}$ \cite{se_analysis2018nips}.
The extrinsic variance and mean of the denoiser are respectively given by
\begin{align}
    v_{\sf B}^{\rm ext} &= \left(\frac{1}{v_{\sf B}^{\rm post}} - \frac{1}{v_{\sf B}^{\rm pri}}\right)^{-1}, \\
    \mathbf{x}_{\sf B}^{\rm ext} &= v_{\sf B}^{\rm ext} \left(\frac{\mathbf{x}_{\sf B}^{\rm post}}{v_{\sf B}^{\rm post}} - \frac{\mathbf{x}_{\sf B}^{\rm pri}}{v_{\sf B}^{\rm pri}} \right).
\end{align}
Finally, the extrinsic output of the MMSE denoiser serves as the input to the linear estimator in the next iteration: $\mathbf{x}_{\sf A}^{\rm pri} = \mathbf{x}_{\sf B}^{\rm ext}$, $v_{\sf A}^{\rm pri} = v_{\sf B}^{\rm ext}$.
The two modules are executed iteratively and empirically converge within $10$ iterations \cite{cai2025score, cai2025massive_connectivity}.

\subsubsection{General Inverse Problem}
The STMP framework \cite{cai2025score} holds great potential for addressing general inverse problems characterized by the model $\mathbf{y} = \mathcal{A}(\mathbf{x}) + \sigma \mathbf{n}$.
Specifically, to cope with the non-linearity in the
measurement model, future work may consider replacing the linear estimator with
model-specific message updates derived from the probabilistic factorization of
the posterior distribution. This will enable the developed algorithm to effectively capture
the structure of the non-linear measurement model while retaining the fast convergence of
score-based Bayesian inference.

\subsection{Scoring MI and KLD}\label{sec:scoreMI}

The MI between the channel input $\mathbf{x}$ and the channel output $\mathbf{y}$ can be expressed as 
\begin{align}\label{eq:MI-log-density}
I(\mathbf{x},\mathbf{y})=\mathbb{E}_{p(\mathbf{x},\mathbf{y})}[\log p(\mathbf{x}|\mathbf{y})-\log p(\mathbf{x})].
\end{align} 
The KLD between $p(\mathbf{x}|\mathbf{y})$ and $  p(\mathbf{x})$ can be expressed as 
\begin{align}\label{eq:KLD-log-density}
{D}_{\rm KL}\left(p(\mathbf{x}|\mathbf{y})||  p(\mathbf{x})\right)=\mathbb{E}_{p(\mathbf{x}|\mathbf{y})}[\log p(\mathbf{x}|\mathbf{y})-\log p(\mathbf{x})].
\end{align} 
To utilize the score-based generative model for MI and KLD evaluations, in the following, we first formulate the log-densities $\log p (\mathbf{x})$ and $\log p (\mathbf{x}|\mathbf{y})$ in terms of score functions. We then show that these score functions can be learned via \ac{DSM}, allowing the MI and KLD evaluations.

\subsubsection{Log-Density Formula}
To enable a score-based characterization of the log-density $\log p (\mathbf{x})$, we begin by introducing a virtual \ac{AWGN} channel, where the input $\mathbf{x}$ is perturbed by Gaussian noise. The output of this virtual channel is
\begin{align}
    \mathbf{x}_t=\mathbf{x}+\sigma_t \mathbf{n},
\end{align}
where $\mathbf{n} \sim ~\mathcal{N}(0,\mathbf{I})$ and $\sigma_t$ controls the noise level.
This virtual channel connects the original data $\mathbf{x}$ to its noise-smoothed version $\mathbf{x}_t$, where we can apply well-established tools like MMSE and score functions, as introduced in Sec.~\ref{subsec:DSM} and Sec.~\ref{sec:scoreMMSE}.  
To further structure this score-based formulation, we relate the unknown true distribution $p(\mathbf{x})$ to a known Gaussian reference $p_G(\mathbf{x}) = \mathcal{N}(\mathbf{0}, \mathbf{I})$, whose log-density, MMSE, and score are all analytically tractable.
Under the true prior $p(\mathbf{x})$ and the Gaussian prior $p_G(\mathbf{x})$, the virtual \ac{AWGN} channel induces the following marginal output distributions:
\begin{align}
p(\mathbf{x}_t)&=\int p(\mathbf{x}_t|\mathbf{x}) p(\mathbf{x}) \mathrm{d} \mathbf{x},\\
p_G(\mathbf{x}_t)&=\int p(\mathbf{x}_t|\mathbf{x}) p_G(\mathbf{x}) \mathrm{d} \mathbf{x}.
\end{align}
Notably, under the Gaussian reference $p_G(\mathbf{x})$, the output distribution $p_G(\mathbf{x}_t)$ remains Gaussian at all noise levels. In contrast, for $p(\mathbf{x})$, at low noise levels ($\sigma_t\to 0$), the output distribution $p(\mathbf{x}_t)$ closely approximates $p(\mathbf{x})$; as the noise level increases towards infinity ($\sigma_t\to \infty$), the noise dominates and $p(\mathbf{x}_t)$ converges to the Gaussian distribution $p_G(\mathbf{x}_t)$. 
By analyzing how the noise-smoothed distributions $p(\mathbf{x}_t)$ and $p_G(\mathbf{x}_t)$ evolve differently as noise scales, we can quantify the discrepancy between $p(\mathbf{x})$ and $p_G(\mathbf{x})$, ultimately yielding a formula for $\log p(\mathbf{x})$ in terms of a known Gaussian term and a score function~\cite{kong2023information,kong2023interpretable}, as detailed below.

This noise-scaling analysis is conceptually linked to the I-MMSE relationship introduced in Sec.~\ref{subsubsec:MI-MMSE}, which reveals that in an AWGN channel, the rate of MI increase as the \ac{SNR} increases (or noise decreases) is equal to half of the MMSE. 
Specifically, for the unknown true distribution $p(\mathbf{x})$, we rewrite the I-MMSE relationship in \eqref{eq:I-MMSE} as 
\begin{align}\label{eq:I-MMSE-noise}
    \frac{\mathrm{d} I(\mathbf{x}, \mathbf{x}_t)}{\mathrm{d}\sigma_t^{-2}} & = \frac{1}{2}{\rm MMSE}(\sigma_t),
\end{align}
where 
\begin{subequations}\label{eq:MI-MMSE-noise}
\begin{align}\label{eq:MI-noise}
I(\mathbf{x}, \mathbf{x}_t)
 &= \mathbb{E}_{p(\mathbf{x})}\!\left[D_{\rm KL}(p(\mathbf{x}_t|\mathbf{x})|| p(\mathbf{x}_t))\right],  
\\ \label{eq:MMSE-noise}
 {\rm MMSE}(\sigma_t) &= \mathbb{E}_{p(\mathbf{x})p(\mathbf{x}_t|\mathbf{x})}\!\left[  \left\| \mathbf{x}-\hat{\mathbf{x}}_{\rm mmse}(\mathbf{x}_t,\sigma_t) \right\| ^2_2 \right] ,  
\end{align}    
\end{subequations}
where $\hat{\mathbf{x}}_{\rm mmse}(\mathbf{x}_t,\sigma_t)=\mathbb{E}_{p(\mathbf{x}|\mathbf{x}_t)}[\mathbf{x}]$ is the MMSE estimator.
This relationship also holds point-wise in $\mathbf{x}$ as~\cite{kong2023information}
\begin{align}\label{eq:pointI-MMSE}
\frac{\mathrm{d} D_{\rm KL}(p(\mathbf{x}_t|\mathbf{x})|| p(\mathbf{x}_t))  }{\mathrm{d}\sigma_t^{-2}}&=\frac{1}{2}  \underbrace{\mathbb{E}_{p(\mathbf{x}_t|\mathbf{x})}[  \left\| \mathbf{x}-\hat{\mathbf{x}}_{\rm mmse}(\mathbf{x}_t,\sigma_t) \right\| ^2_2]}_{{\rm mmse}(\mathbf{x},\sigma_t)},
\end{align}
where $\mathrm{mmse}(\mathbf{x}, \sigma_t)$ is the conditional MMSE given $\mathbf{x}$.
To compare with a known Gaussian baseline, we replace $p(\mathbf{x})$ with $p_G(\mathbf{x})$ in \eqref{eq:pointI-MMSE}, which gives 
\begin{align}\label{eq:pointI-MMSE-gaussian}
\frac{\mathrm{d} D_{\rm KL}(p(\mathbf{x}_t|\mathbf{x})|| p_G(\mathbf{x}_t))  }{\mathrm{d}\sigma_t^{-2}}&=\frac{1}{2}  {\rm mmse}_G(\mathbf{x},\sigma_t),
\end{align}
where ${\rm mmse}_G(\mathbf{x},\sigma_t)$ has a closed-form expression given by~\cite{kong2023information} 
\begin{align}\label{eq:mmseG}
 {\rm mmse}_G(\mathbf{x},\sigma_t)&= \frac{\left\| \mathbf{x} \right\|_2^2+D\sigma_t^{-2}}{(1+\sigma_t^{-2})^2}, \text{ for } \mathbf{x}\in\mathbb{R}^D.
\end{align}

To quantify the discrepancy between how $p(\mathbf{x})$ and $p_G(\mathbf{x})$ behave when passed through the same AWGN channel, we define
\begin{align}
\begin{split}
  & f(\mathbf{x},\sigma_t)= D_{\rm KL}(p(\mathbf{x}_t|\mathbf{x})|| p_G(\mathbf{x}_t)) -  D_{\rm KL}(p(\mathbf{x}_t|\mathbf{x})|| p(\mathbf{x}_t)),
\end{split}
\end{align}
which satisfies $\mathbb{E}_{p(\mathbf{x})}[f(\mathbf{x},\sigma_t)]=D_{\rm KL}(p(\mathbf{x}_t)|| p_G(\mathbf{x}_t)) $. This measures how far the output distribution $p(\mathbf{x}_t)$, generated from $p(\mathbf{x})$, deviates from the Gaussian output distribution $p_G(\mathbf{x}_t)$.
As mentioned before, when the noisy power becomes infinity, the output distributions are almost Gaussian regardless of the input distributions; when there is no noise, the distance between the output distributions equals the distance between the input distributions, i.e.,~\cite{kong2023information} 
\begin{align}
\lim_{\sigma_t\to 0} f(\mathbf{x},\sigma_t) & = \log \frac{p(\mathbf{x})}{p_G(\mathbf{x})},\\ \lim_{\sigma_t\to \infty} f(\mathbf{x},\sigma_t) &= 0.
\end{align} 
Applying Thermodynamic integration~\cite{masrani2019thermodynamic,brekelmans2020all}, we have
\begin{align}
\begin{split}
  \int_{0}^{\infty} {\mathrm{d}\sigma_t}\frac{\mathrm{d} }{\mathrm{d}\sigma_t} f(\mathbf{x},\sigma_t)&=\lim_{\sigma_t\to \infty} f(\mathbf{x},\sigma_t)-\lim_{\sigma_t\to 0} f(\mathbf{x},\sigma_t)\\&=-\log \frac{p(\mathbf{x})}{p_G(\mathbf{x})}. 
\end{split}
\end{align}
This can be rewritten as
\begin{align} \label{eq:-logpx1}
          &-\log p(\mathbf{x})
          =-\log p_G(\mathbf{x}) +\int_{0}^{\infty} {\mathrm{d}\sigma_t^{-2}}\frac{\mathrm{d} }{\mathrm{d}\sigma_t^{-2}} f(\mathbf{x},\sigma_t)
          \nonumber\\&\overset{(a)}{=} \! -\log p_G(\mathbf{x})\!+\!\frac{1}{2} \!\int_{0}^{\infty}  \!\!\!\!\big ( {\rm mmse}_G(\mathbf{x},\sigma_t)- {\rm mmse}(\mathbf{x},\sigma_t) \big ) \mathrm{d} \sigma_t^{-2},
    \end{align}
    where (a) applies the point-wise I-MMSE relationship in \eqref{eq:pointI-MMSE} and \eqref{eq:pointI-MMSE-gaussian}.  
This expression decomposes the log-density of an unknown true distribution into two parts: the log-density of a known Gaussian distribution and a correction term that integrates the MMSE gap over noise levels.
Moreover, we can explicitly write the log-density of $p_G(\mathbf{x}) = \mathcal{N}(\mathbf{0}, \mathbf{I})$ as
\begin{align}\label{eq:log-pG}
    \log p_G(\mathbf{x})=-\frac{D}{2}\log 2\pi -\frac{1}{2} \left\| \mathbf{x} \right\|_2^2.
\end{align}
Combining \eqref{eq:log-pG} with \eqref{eq:mmseG}, we can simplify \eqref{eq:-logpx1} to 
\begin{align} \label{eq:logpx-mmse}
          &-\log p(\mathbf{x})= -\frac{1}{2} \int_{0}^{\infty} {\rm mmse}(\mathbf{x},\sigma_t) \mathrm{d} \sigma_t^{-2} + {C_G}, 
    \end{align}
where $C_G$ is a constant that depends only on the reference Gaussian distribution.  

Eq.~\eqref{eq:logpx-mmse} reveals that the log-density of $p(\mathbf{x})$ can be characterized by an MMSE estimator under a virtual AWGN channel.
The MMSE estimator can be connected to the score function $\nabla_{\mathbf{x}_t} \log p(\mathbf{x}_t)$ using Tweedie's formula~\cite{robbins1992empirical}: 
\begin{align}
\hat{\mathbf{x}}_{\rm mmse}(\mathbf{x}_t,\sigma_t)=\mathbb{E}_{p(\mathbf{x}|\mathbf{x}_t)}[\mathbf{x}]=\mathbf{x}_t+\sigma_t^2\nabla_{\mathbf{x}_t} \log p(\mathbf{x}_t).
\end{align}
The corresponding MMSE is
\begin{align}\label{eq:mmse-denoiser}
    {\rm mmse}(\mathbf{x},\sigma_t)&= \mathbb{E}_{p(\mathbf{x}_t|\mathbf{x})}[ || \mathbf{x}-\hat{\mathbf{x}}_{\rm mmse}(\mathbf{x}_t,\sigma_t)|| ^2_2] \nonumber\\&= \mathbb{E}_{p(\mathbf{n})}[ || \sigma_t\mathbf{n}+\sigma_t^2\nabla_{\mathbf{x}_t} \log p(\mathbf{x}_t)|| ^2_2].
\end{align}
Substituting \eqref{eq:mmse-denoiser} into \eqref{eq:logpx-mmse} yields the score-based log-density formula:
\begin{align}
    \label{eq:logpx-score}
            &-\log p(\mathbf{x})
            \nonumber\\&=-\frac{1}{2} \int_{0}^{\infty} \mathbb{E}_{p(\mathbf{n})}[  \left\| \sigma_t\mathbf{n}+\sigma_t^2\nabla_{\mathbf{x}_t} \log p(\mathbf{x}_t) \right\| ^2_2] \mathrm{d} \sigma_t^{-2} + {C_G}
            \nonumber\\&\overset{(a)}{=} \int_{0}^{\infty} \!\!\sigma_t^{-1}\mathbb{E}_{p(\mathbf{n})}[  \left\| \mathbf{n}+\sigma_t\nabla_{\mathbf{x}_t} \log p(\mathbf{x}_t) \right\|  ^2_2] \mathrm{d} \sigma_t + {C_G},
\end{align} 
where (a) is from $\mathrm{d}\sigma_t^{-2}=-2\sigma_t^{-3}\mathrm{d}\sigma_t$.
The conditional log-density can be expressed in a similar form as~\cite{kong2023interpretable}
\begin{align}
\label{eq:logp(x|y)-score} 
            &-\log p(\mathbf{x}|\mathbf{y}) \nonumber\\& =  \int_0^{\infty}  \!\!\sigma_t^{-1} \mathbb{E}_{p(\mathbf{n})}[ \left\| \mathbf{n}+ \sigma_t\nabla_{\mathbf{x}_t} \log p(\mathbf{x}_t|\mathbf{y}) \right\| _2^2] \mathrm{d} \sigma_t+{C_G}.
    \end{align}

\subsubsection{Scoring MI and KLD With DSM}

With \eqref{eq:logpx-score} and \eqref{eq:logp(x|y)-score},   we can reformulate $I(\mathbf{x},\mathbf{y})$ in \eqref{eq:MI-log-density} as
\begin{align}\label{eq:MI-score}
&I(\mathbf{x},\mathbf{y})
=\mathbb{E}_{p(\mathbf{x},\mathbf{y})}\bigg[\int_0^{\infty} \sigma_t^{-1}\mathbb{E}_{p(\mathbf{n})}\big[   \left\| \mathbf{n}+\sigma_t \nabla_{\mathbf{x}_t} \log p(\mathbf{x}_t) \right\| _2^2 
\nonumber\\&  \qquad\qquad -  \left\| \mathbf{n}+\sigma_t\nabla_{\mathbf{x}_t} \log p(\mathbf{x}_t|\mathbf{y})\right\|_2^2\big] \mathrm{d} \sigma_t 
\bigg].
    \end{align}
To estimate the smoothed unconditional and conditional scores, i.e.,  $\nabla_{\mathbf{x}_t} \log p(\mathbf{x}_t)$ and  $\nabla_{\mathbf{x}_t} \log p(\mathbf{x}_t|\mathbf{y})$ in \eqref{eq:MI-score}, we apply the DSM-based posterior approach in Sec.~\ref{sec:cond-DSM}. We then use the learned unconditional and conditional scores $\mathbf{s}_{\bm \phi_{\rm DSM}}(\mathbf{x}_t ,\sigma_t)$ and $\mathbf{s}_{\bm \phi_{\rm DSM}}(\mathbf{x}_t,\sigma_t|\mathbf{y})$ for approximating $\nabla_{\mathbf{x}_t } \log p(\mathbf{x}_t)$ and $ \nabla_{\mathbf{x}_t} \log p(\mathbf{x}_t|\mathbf{y})$, respectively.
The MI $I(\mathbf{x},\mathbf{y})$ in \eqref{eq:MI-score} can be evaluated by
\begin{align}\label{eq:MI-DSM}
\hat{I}(\mathbf{x},\mathbf{y})
&=\mathbb{E}_{p(\mathbf{x},\mathbf{y})}\bigg[\int_0^{\infty} \sigma_t^{-1}\mathbb{E}_{p(\mathbf{n})}\big[ 
\left\| \mathbf{n}+\sigma_t\mathbf{s}_{\bm \phi_{\rm DSM}}(\mathbf{x}_t,\sigma_t)  \right\|_2^2
\nonumber\\&  \quad - \left\| \mathbf{n}+\sigma_t \mathbf{s}_{\bm \phi_{\rm DSM}}(\mathbf{x}_t,\sigma_t|\mathbf{y})  \right\|_2^2
\big] \mathrm{d} \sigma_t
\bigg].
    \end{align}
Note that the expectation operations $\mathbb{E}_{p(\mathbf{n})}[\cdot]$ and $\mathbb{E}_{p(\mathbf{x},\mathbf{y})}[\cdot]$ in \eqref{eq:MI-DSM} can be approximated by Monte Carlo methods using data samples of $\mathbf{n}$, $\mathbf{x}$, and $\mathbf{y}$. Moreover, the integration over $\sigma_t$ can be conducted by importance sampling with finite noise levels $\{\sigma_t\}_{t=1}^T$~\cite{kong2023information}.

Based on \eqref{eq:KLD-log-density}, we extend the framework of scoring MI to KLD by
\begin{align}\label{eq:KLD-score}
&\hat{D}_{\rm KL}\left(p(\mathbf{x}|\mathbf{y})||  p(\mathbf{x})\right) 
\nonumber
\\&=\mathbb{E}_{p(\mathbf{x}|\mathbf{y})}\bigg[\int_0^{\infty}  \sigma_t^{-1}\mathbb{E}_{p(\mathbf{n})}\big[ 
 \left\| \mathbf{n}+\sigma_t\mathbf{s}_{\bm \phi_{\rm DSM}}(\mathbf{x}_t,\sigma_t) \right\|_2^2
\nonumber\\&  \quad - \left\| \mathbf{n}+\sigma_t \mathbf{s}_{\bm \phi_{\rm DSM}}(\mathbf{x}_t,\sigma_t|\mathbf{y})  \right\| _2^2
\big] \mathrm{d} \sigma_t
\bigg].
    \end{align}

\section{Applications}\label{sec:Sapplication}
In this section, we apply the scoring ISAC framework introduced in Sec.~\ref{sec:Score-ISAC} to evaluate KLD for target detection and BCRB for target localization. 
To validate the score-based approach, 
we generate i.i.d. data samples from a known statistical model with analytically tractable structure.
This allows closed-form expressions for KLD or BCRB to serve as ground truth benchmarks. 
Comparing score-based estimates with these classical model-based analytical results directly assesses the applicability of the scoring ISAC framework.
This proof of concept lays the groundwork for future applications of the framework to more complex and realistic ISAC scenarios.

\begin{figure}
    \centering
    \includegraphics[width=0.9\linewidth]{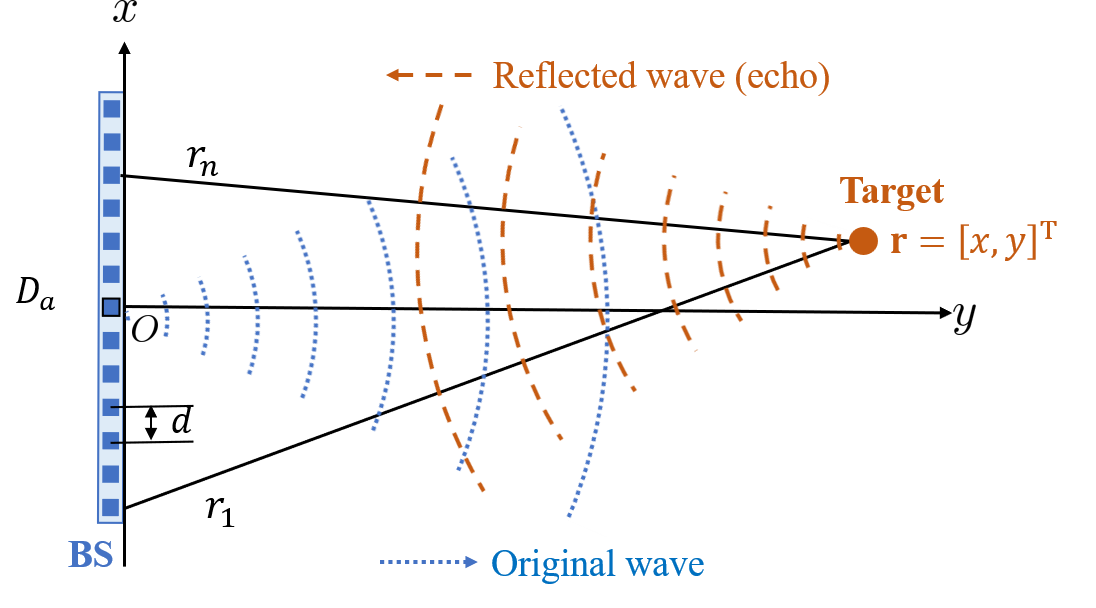}
    \caption{Target detection and localization in the near-field region.}
    \label{fig:NFsensing}
\end{figure}

\subsection{Channel Model}

Consider a narrowband sensing system, where a transmitter and a receiver are co-located at a \ac{BS}~\cite{wang2024near}, as shown in Fig.~\ref{fig:NFsensing}. The BS is equipped with a \ac{ULA} with $N$ antennas. The antenna spacing is $d=D_a/(N-1)$, where $D_a$ is the antenna aperture. 
The ULA lies along the x-axis, and the origin of the coordinate system is set at the center of the ULA. 
We denote the coordinate of the $n$-th antenna as $\mathbf{r}^a_n=[\delta_n d,0]^\mathsf{T}$, where $\delta_n=n-1-\frac{N-1}{2}$, $n\in\{1,...,N\}$.
Assume a sensing target is located at $\mathbf{r}=[x,y]^\mathsf{T}$ in the near-field region of the BS with a \ac{LoS} link. 
The round-trip near-field sensing channel can be modeled as~\cite{wang2023near} 
\begin{align}\label{eq:Hsensing}
\mathbf{H}_{\rm s}=\beta \mathbf{a}(x,y) \mathbf{a}(x,y)^\mathsf{T},
\end{align}
where $\beta$ is the channel coefficient, and $\mathbf{a}(x,y)\in\mathbb{C}^N$ is the near-field array response vector. Specifically, the $n$-th entry of $\mathbf{a}(x,y)$ is given by
\begin{align}
    [\mathbf{a}(x,y)]_n=\frac{1}{r_n}\exp\left(-j\frac{2\pi}{\lambda} r_{n}\right),
\end{align}
where $r_{n}= \left\| \mathbf{r}-\mathbf{r}^a_{n} \right\|=\sqrt{x^2+y^2-2\delta_n d x+ (\delta_n d)^2}$ is the distance between the $n$-th antenna and the target, and $\lambda$ is the wavelength associated with the carrier frequency $f$. 
According to the radar range equation, the channel coefficient can be modeled as $\beta=\beta_0\gamma$~\cite{luo2024integrated,wang2024near}, where $\beta_0=\sqrt{\frac{\lambda^2}{(4\pi)^3}}$ and $\gamma$ is the \ac{RCS} of the target. The \ac{RCS} $\gamma$ characterizes the reflectivity of the target, quantifying how much incident energy is reflected back toward the receiver. It depends on various factors, such as the target's size, shape, material, and orientation~\cite{RCS}.
Under the Swerling I model~\cite{luo2024integrated}, $\gamma$ can be modeled as a random variable with an exponential prior distribution $\gamma\sim\rm Exp(\gamma_0)$, i.e., $p(\gamma)=\frac{1}{\gamma_0}\exp(-\frac{\gamma}{\gamma_0})$.

In the following target detection and localization problems, we assume the sensing process collects $K$ i.i.d. measurements within a coherent time block, during which the sensing target parameters remain fixed. The transmitted sensing signal at each measurement snapshot is $\mathbf{x}$, 
where $\mathbb{E}[ \left\| \mathbf{x} \right\|_2^2]=P_{\rm t}$  and $P_{\rm t}$ is the transmitted power. 
The $k$-th received echo is denoted as $\mathbf{y}_{\rm s}[k]$ and  {$\mathbf{z}_{\rm s}[k]\sim\mathcal{CN}(\mathbf{0},\sigma_{\rm s}^2\mathbf{I})$ denotes the corresponding noise. The default values of system parameters are given in Table~\ref{tab:simulation}.

 \begin{table}[t!]\caption{Default values of system parameters.}
\vspace{-3mm}
\centering
\begin{center}
{\linespread{0.95}
\renewcommand{\arraystretch}{1.2} \small
    \begin{tabular}{ | {c} | {c} | {c} | }
    \hline
        \hline
    {Parameters} & {Descriptions} & {Default Values} \\ \hline
    $N$ & Antenna number & $64$ \\ \hline
    
    $K$ & Number of measurement & $4$  \\ \hline
    $f$ & Carrier frequency & $28~\rm GHz$  \\ \hline
    $x,y$ & Target location & $20~\rm m, 20~\rm m$ \\ \hline
    $P_{\rm t}$ & Transmit power & $20~\rm dBm$ \\ \hline
    $\sigma_{\rm s}^2$ & Noise power & $-60~\rm dBm$  \\ \hline
    $D_a$ & Antenna aperture & $0.5~\rm m$  \\ \hline
    \end{tabular}}\color{black}
\end{center}
\label{tab:simulation}
\vspace{-2mm}
\end{table}

\subsection{KLD of Target Detection}
We consider a detection task that determines whether a target is present or absent at $\mathbf{r}=[x,y]^\mathsf{T}$. 
This can be formulated as a binary hypothesis testing problem:   
\begin{align}\label{eq:y_k_complex}
   \mathbf{y}_{\rm s} [k]= \begin{cases}
        \mathcal{H}_1: \mathbf{H}_{\rm s}\mathbf{x}+\mathbf{z}_{\rm s}[k], k=1,..., K,\\
        \mathcal{H}_0: \mathbf{z}_{\rm s},k=1,..., K,\\
    \end{cases}
\end{align}
where $\mathbf{H}_{\rm s}$ is given in \eqref{eq:Hsensing}, $\mathcal{H}_0$ and $\mathcal{H}_1$ correspond to the hypotheses of target absence and presence, respectively. 
The objective is to infer the correct hypothesis based on the received echo signals over $K$ measurements, i.e., $\mathbf{Y}_{\rm s}=[\mathbf{y}_{\rm s}[1],...,\mathbf{y}_{\rm s}[K]]$.
As introduced in Sec.~\ref{sec:detection}, the KLD $D_{\rm KL}\left( p(\mathbf{Y}_{\rm s}|\mathcal{H}_0) ||  p(\mathbf{Y}_{\rm s}|\mathcal{H}_1) \right)$ can be used to measure the detection performance. In the following, we present the analytical and score-based methods for computing KLD.
We consider two cases: (i) a special case where the RCS is known and fixed as $\gamma = 1$, and (ii) a more general case where $\gamma$ is unknown and modeled as a random variable with $\gamma \sim \mathrm{Exp}(1)$.

\subsubsection{Known Reflection}
Under the special case of $\gamma=1$, the likelihoods under the hypotheses of target absence and presence are given by
\begin{align}
   & p(\mathbf{Y}_{\rm s}|\mathcal{H}_0) =\prod_{k=1}^{K} p(\mathbf{y}_{\rm s}[k]|\mathcal{H}_0) = \prod_{k=1}^{K} \mathcal{CN}(\mathbf{0},\sigma_{\rm s}^2\mathbf{I}), \label{eq:pdf_y_H0}
    \\&  p(\mathbf{Y}_{\rm s}|\mathcal{H}_1) =\prod_{k=1}^{K} p(\mathbf{y}_{\rm s}[k]|\mathcal{H}_1)  = \prod_{k=1}^{K} \mathcal{CN}(\mathbf{H}_{\rm s}\mathbf{x},\sigma_{\rm s}^2\mathbf{I}).\label{eq:pdf_y_H1}
\end{align} 
Note that the above random variables are complex-valued. For the subsequent analysis, we reformulate \eqref{eq:y_k_complex} as
\begin{align}\label{eq:y_k_real}
   {\tilde{\mathbf{y}}_{\rm s}} [k]= \begin{cases}
        \mathcal{H}_1: \tilde{\mathbf{h}}+\tilde{\mathbf{z}}_{\rm s}[k], k=1,..., K,\\
        \mathcal{H}_0: \tilde{\mathbf{z}}_{\rm s}[k],k=1,..., K,\\
    \end{cases}
\end{align}
where the real-valued formulation of $\mathbf{y}_{\rm s}[k]\in\mathbb{C}^N$ is denoted by $\tilde{\mathbf{y}}_{\rm s}[k]=[{\rm Re}\{ {\mathbf{y}}_{\rm s}[k]^{\mathsf{T}}\},{\rm Im}\{ {\mathbf{y}}_{\rm s}[k]^{\mathsf{T}}\}]^{\mathsf{T}}\in\mathbb{R}^{2N}$, and $\tilde{\mathbf{h}}$ is the real-valued formulation of $\mathbf{h}=\mathbf{H}_{\rm s}\mathbf{x}$, and $\tilde{\mathbf{z}}_{\rm s}[k]\sim\mathcal{N}(\mathbf{0},\tilde{\sigma}_{\rm s}^2\mathbf{I})$ with $\tilde{\sigma}_{\rm s}^2=\frac{ \sigma_{\rm s}^2}{2}$. Then, we have 
\begin{subequations}
\begin{align}\label{eq:yk_real}
    &p(\tilde{\mathbf{y}}_{\rm s}[k]|\mathcal{H}_0)=\mathcal{N}(\mathbf{0},\tilde{\sigma}_{\rm s}^2\mathbf{I}) \text{ and }p(\tilde{\mathbf{y}}_{\rm s}[k]|\mathcal{H}_1)=\mathcal{N}(\tilde{\mathbf{h}},\tilde{\sigma}_{\rm s}^2\mathbf{I}).
\end{align}     
\end{subequations}
We denote the KLD between $p(\mathbf{Y}_{\rm s}|\mathcal{H}_0)$ and $p(\mathbf{Y}_{\rm s}|\mathcal{H}_1)$ as $D_{\rm KL}^{\gamma=1}$, which has a closed-form expression given by
\begin{subequations}\label{eq:analytical_KLD}
\begin{align}
    D_{\rm KL}^{\gamma=1} 
    & =\sum_{k=1}^{K}D_{\rm KL} (p(\tilde{\mathbf{y}}_{\rm s}[k]|\mathcal{H}_0) ||p(\tilde{\mathbf{y}}_{\rm s}[k]|\mathcal{H}_1))  \label{eq:KLD-k} 
     \\&=K\frac{\left\| \mathbf{H}_{\rm s}\mathbf{x} \right\|^2_2}{\sigma_{\rm s}^2}.\label{eq:analy_KLD}
\end{align}    
\end{subequations}

To demonstrate the correlation between the KLD $D_{\rm KL}^{\gamma=1}$ and the detection probability $P_{\rm d}$, as well as to illustrate the feasibility of estimating KLD via score-based techniques, we present the results in Fig.~\ref{fig:KLD_vs_P_D}. In this experiment, the transmit signal $\mathbf{x}$ is chosen to be aligned with the {right singular vector} corresponding to the sole nonzero {singular value} of $\mathbf{H}_{\rm s}$. The analytical KLD and the learned KLD are computed using \eqref{eq:analytical_KLD} and \eqref{eq:KLD-score}, respectively. In particular, we train a single-layer linear neural network without bias and activation using DSM to learn the score function of the Gaussian distribution $\mathbf{u} \sim \mathcal{N}(\mathbf{0}, \mathbf{I})$, scaled by $(1 + s_t^2)$, where $s_t > 0$. The noise smoothed data is {$\mathbf{u}_t=\mathbf{u}+s_t\mathbf{n}$, where $\mathbf{n}\sim \mathcal{N}(\mathbf{0}, \mathbf{I})$}. The model is optimized using Adam with a learning rate of $0.0001$ for $300{,}000$ iterations on a training set of size $10{,}000$, with a batch size of $128$. Exponential moving average (EMA) with a decay rate of $0.999$ is applied during training. {We denote the learned noisy smoothed score, obtained by dividing the neural network output by $(1 + s_t^2)$, as $\mathbf{s}_{\bm \phi_{\rm DSM}}(\mathbf{u}_t,s_t)$. 
Then, the smoothed scores of $\tilde{\mathbf{y}} \sim p(\tilde{\mathbf{y}}_{\rm s}[k]|\mathcal{H}_0)$ and $\tilde{\mathbf{y}} \sim p(\tilde{\mathbf{y}}_{\rm s}[k]|\mathcal{H}_1)$ in \eqref{eq:yk_real} are given by {$\frac{1}{\tilde{\sigma}_{\rm s}}\mathbf{s}_{\bm \phi_{\rm DSM}}\left(\frac{\tilde{\mathbf{y}} +\sigma_t \mathbf{n}}{\tilde{\sigma}_{\rm s}},\frac{\sigma_t}{\tilde{\sigma}_{\rm s}}\right)$} 
and {$\frac{1}{\tilde{\sigma}_{\rm s}}\mathbf{s}_{\bm \phi_{\rm DSM}}\left(\frac{\tilde{\mathbf{y}} +\sigma_t \mathbf{n} - \tilde{\mathbf{h}}}{\tilde{\sigma}_{\rm s}},\frac{\sigma_t}{\tilde{\sigma}_{\rm s}}\right)$}, respectively.}
Then, with \eqref{eq:KLD-score} and \eqref{eq:KLD-k}, we obtain the learned KLD.
As for the detection probability $P_{\rm d}$, it is calculated analytically by \eqref{eq:pd_pfa}, where the false alarm probability is set to $P_{\rm fa} = 0.1$. As shown in Fig.~\ref{fig:KLD_vs_P_D}, the detection probability $P_{\rm d}$ increases with the KLD, in line with the discussion in Section \ref{sec:detection}. Moreover, the KLDs estimated via score-based techniques closely match their analytical counterparts, providing preliminary evidence of these techniques.

\begin{figure}
\vspace{-3mm}
    \centering
    \includegraphics[width=0.85\linewidth]{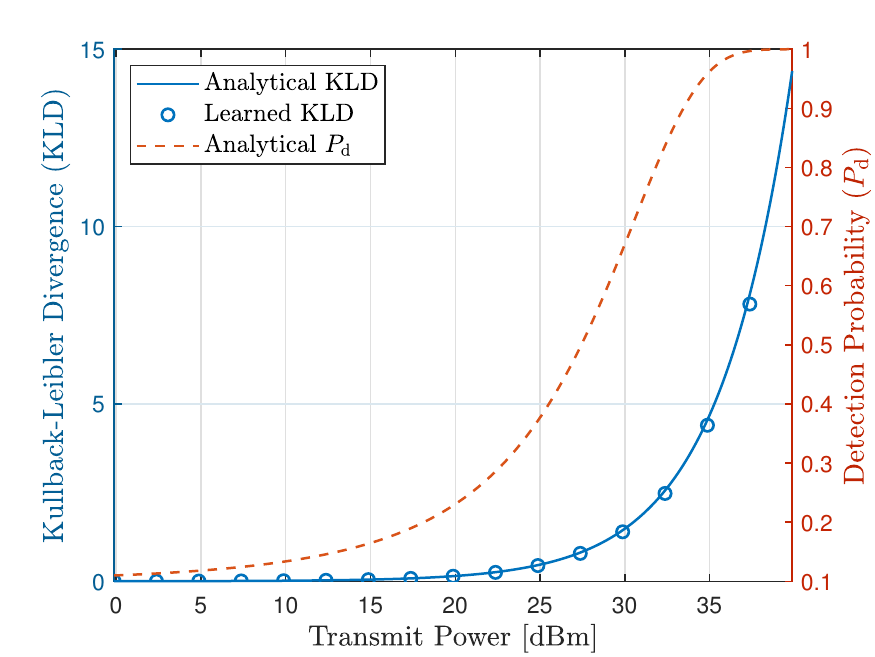}
    \caption{Target detection performance with $\gamma=1$.}
    \label{fig:KLD_vs_P_D}
\end{figure}

\begin{figure}
    \centering
    \includegraphics[width=0.85\linewidth]{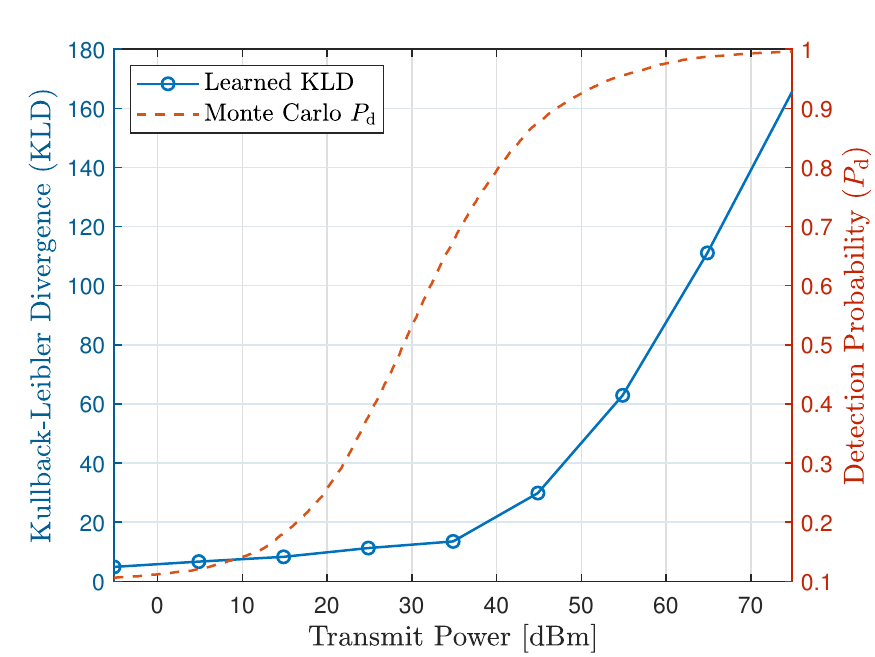}
    \caption{Target detection performance with $\gamma\sim{\rm Exp}(1)$.}
    \label{fig:nonGaussianKLD_vs_P_D}
\end{figure}

\subsubsection{Unknown Reflection}
 
We now consider the general case where $\gamma\sim\rm Exp(1)$. The likelihoods under the two hypotheses are given by
\begin{align}
    p(\mathbf{Y}_{\rm s}|\mathcal{H}_0) 
    =& \prod_{k=1}^{K} \mathcal{CN}(\mathbf{0},\sigma_{\rm s}^2\mathbf{I}), \label{eq:nonG_pdf_y_H0}
    \\ p(\mathbf{Y}_{\rm s}|\mathcal{H}_1)  
    =& \prod_{k=1}^{K} \mathbb{E}_{\gamma \sim \mathrm{Exp}(1)} \left\{\mathcal{CN}(\mathbf{H}_{\rm s}\mathbf{x},\sigma_{\rm s}^2\mathbf{I})\right\}.\label{eq:nonG_pdf_y_H1}
\end{align} 
In this case, neither the KLD $D_{\rm KL}$ nor the detection probability $P_{\rm d}$ admits a closed-form expression. Instead, we estimate $P_{\rm d}$ of a practical detection scheme via Monte Carlo and estimate $D_{\rm KL}$ using score-based techniques. The correlation between $P_{\rm d}$ and $D_{\rm KL}$ is then visualized through plotting. 
 
Specifically, we still consider that the transmit signal $\mathbf{x}$ is aligned with the {right singular vector} corresponding to the sole nonzero {singular value} of $\mathbf{H}_{\rm s}$. The considered detection scheme is the optimal scheme for the case $\gamma = 1$, which is well-established in the literature (e.g. \cite{cover1999elements}). Naturally, as the detection scheme does not take into account the randomness of $\gamma$, this scheme is suboptimal in general. {Under the above setting, we estimate $P_{\rm d}$ by computing the frequency of successful detections with the false alarm probability being approximately $0.1$.} The KLD is computed using \eqref{eq:KLD-score} with the score functions approximated by neural networks trained via DSM. For training details, we adopt the class-conditional score network architecture proposed in \cite{ho2022classifier}, and train it for $1{,}000$ epochs with a batch size of $128$ using the Adam optimizer with a learning rate of $0.0001$ on a synthetic dataset comprising $10$ classes. Class $1$ corresponds to the null hypothesis $\mathcal{H}_0$, while the remaining $9$ classes correspond to $9$ different \ac{SNR} levels under the hypothesis $\mathcal{H}_1$. Each class contains $10{,}000$ training samples. The results are shown in Fig.~\ref{fig:nonGaussianKLD_vs_P_D}, where a clear positive correlation between $P_{\rm d}$ and the KLD can be observed. {Comparing Fig.~\ref{fig:KLD_vs_P_D} and Fig.~\ref{fig:nonGaussianKLD_vs_P_D}, we observe that, under the same transmit power, the KLD is higher and the detection probability $P_{\rm d}$ is lower in Fig.~\ref{fig:nonGaussianKLD_vs_P_D}. The increase in KLD is due to the additional randomness introduced by $\gamma$, which, despite satisfying $\mathbb{E}[\gamma] = 1$, enlarges the ``discrepancy'' between the two distributions involved in the KLD computation. The reduction in $P_{\rm d}$ in Fig.~\ref{fig:nonGaussianKLD_vs_P_D}, on the other hand, results from the use of the detection scheme that is suboptimal for its considered system model, whereas the underlying detection scheme in Fig.~\ref{fig:KLD_vs_P_D} is optimal for its corresponding model.}

\subsection{BCRB of Target Localization}
The goal of target localization is to estimate the target position $\mathbf{r}=[x,y]^\mathsf{T}$ based on the observed echoes $\{\mathbf{y}_s[k]\}_{k=1}^K$ and prior information about $\mathbf{r}$.
Specifically, the sensing receiver at the BS knows the transmitted signal $\mathbf{x}$ and attempts to infer the sensing channel $\mathbf{H}_{\rm s}$, which embeds the information about the target location $\mathbf{r}$ and the RCS coefficient $\gamma$. 
The observation at the $k$-th snapshot follows the measurement model: 
\begin{align}\label{eq:fk}
\mathbf{y}_s[k]=\bm{f}(\mathbf{r},\gamma)+\mathbf{z}_s[k],
\end{align}
where $\bm{f}(\mathbf{r},\gamma)=\mathbf{H}_{\rm s}\mathbf{x}$ as defined in \eqref{eq:Hsensing}.
Moreover, we assume a Gaussian prior for the target location $p(\mathbf{r})=\mathcal{N}(\bm \mu_{\rm r},\sigma_{\rm r}^2\mathbf{I})$, i.e., the target is likely positioned around $\mu_{\rm r}$ with uncertainty.

As introduced in Sec.~\ref {sec:estimation}, the BCRB matrix, i.e., the inverse of the Bayesian FIM, provides a fundamental lower bound on the estimation error of the target location.  
In the following, we evaluate the BCRB for target localization under the cases of $\gamma = 1$ and $\gamma \sim \mathrm{Exp}(1)$, respectively,
using both analytical and score-based approaches to compute the Bayesian FIM.

\subsubsection{Known Reflection}\label{subsec:BCRBdet}
Assume the RCS is known and fixed at $\gamma=1$.
The unknown parameter is $\mathbf{r}\in \mathbb{R}^2$ and the measurement operator ${\bm f} (\mathbf{r},\gamma)$ in \eqref{eq:fk} is simplified as ${\bm f} (\mathbf{r})$. 
The prior score and prior FIM with respect to $\mathbf{r}$ can be analytically calculated as
\begin{align}
   & \nabla_{\mathbf{r}} \log p(\mathbf{r}) = -\frac{\mathbf{r} - \bm\mu_{\rm r}}{\sigma_{\rm r}^2},
    \\& \mathbf{J}_{{\rm p},\mathbf{r}} = \mathbb{E}_{p(\mathbf{r})} \left[ \left( \nabla_{\mathbf{r}} \log p(\mathbf{r}) \right) \left( \nabla_{\mathbf{r}} \log p(\mathbf{r}) \right)^\mathsf{T} \right] = \frac{1}{\sigma_{\rm r}^2} \mathbf{I}.
\end{align}
Moreover, the FIM related to the likelihood can be analytically expressed as
\begin{align}\label{eq:analyCRB}
    \mathbf{J}(\mathbf{r})= \sum_{k=1}^{K}\frac{2}{\sigma^2} \Re \left\{ \left( \nabla_{\mathbf{r}} \bm{f}(\mathbf{r}) \right)^\mathsf{H}  \left( \nabla_{\mathbf{r}} \bm{f}(\mathbf{r}) \right)\right\},
\end{align}
where $\nabla_{\mathbf{r}} \bm{f}(\mathbf{r})=\frac{\partial \mathbf{H}_{\rm s} \mathbf{x}}{\partial \mathbf{r}}\in\mathbb{C}^{N\times 2}$.
Following \eqref{eq:BFIM-Bayes}-\eqref{eq:FIM_D}, the analytical Bayesian FIM is given by
\begin{align}\label{eq:analyBCRB}
    \mathbf{J}_{{\rm b},\mathbf{r}} = 
    \mathbf{J}_{{\rm p},\mathbf{r}} + \mathbf{J}_{{\rm d},\mathbf{r}} = 
    \mathbf{J}_{{\rm p},\mathbf{r}}+\mathbb{E}_{p(\mathbf{r})}\mathbf[\mathbf{J}(\mathbf{r})].
\end{align}
Here, the data FIM $\mathbf{J}_{{\rm d}}=\mathbb{E}_{p(\mathbf{r})}\mathbf[\mathbf{J}(\mathbf{r})]$ is generally evaluated by Monte-Carlo methods with $\mathbf{J}_{{\rm d},\mathbf{r}}=\frac{1}{M}\sum_{m=1}^{M}\mathbf{J}(\mathbf{r}_m)$, where $\mathbf{r}_m$ is sampled from $p(\mathbf{r})$. 

In the case of $\gamma=1$, we denote the analytical error bounds of the target localization as
\begin{align}\nonumber
    \mathcal{E}_{\rm BCRB}^{\gamma=1}={\rm Tr}(\mathbf{J}_{{\rm b},\mathbf{r}}^{-1}) \text{  and  } \mathcal{E}_{\rm CRB}^{\gamma=1}={\rm Tr}(\mathbf{J}_{{\rm d},\mathbf{r}}^{-1}),
\end{align}
where $\mathcal{E}_{\rm BCRB}^{\gamma=1}$ represents the analytical \ac{BCRB} that incorporates both the prior and data information and $\mathcal{E}_{\rm CRB}^{\gamma=1}$ represents the analytical \ac{CRB} (without the prior information). 
Consider the scenario where the likelihood is known while the prior distribution is unavailable. Leveraging the ISM-based measurement-prior approach introduced in Sec.~\ref{subsec:ISM-meas-prior}, the learned BCRB is
\begin{align}
    \hat{\mathcal{E}}_{\rm BCRB}^{\gamma=1}={\rm Tr}((\hat{\mathbf{J}}_{{\rm p},\mathbf{r}}+{\mathbf{J}}_{{\rm d},\mathbf{r}})^{-1}), 
\end{align}
where $\hat{\mathbf{J}}_{{\rm p},\mathbf{r}}=\frac{1}{M}\sum_{m=1}^{M}\left[ (\mathbf{s}_{\bm \phi_{\rm ISM}}(\mathbf{r}_m)) (\mathbf{s}_{\bm \phi_{\rm ISM}}(\mathbf{r}_m))^\mathsf{T} \right]$ and $\mathbf{s}_{\bm \phi_{\rm ISM}}(\mathbf{r})$ is the learned prior score.

In Fig.\ref{fig:BCRB_vs_CRB}(a), we compare the analytical BCRB, analytical CRB, and learned BCRB in the case of $\gamma=1$. The prior distribution of the target location is specified by  $\bm\mu_{\rm r} = [20, 20]^\mathsf{T}$ and  $\sigma_{\rm r}^2 = 5$. The transmitted signal is set as $\mathbf{x}\sim \mathcal{CN}(\mathbf{0}, \frac{P_{\rm t}}{N} \mathbf{I})$, which is initially randomly generated and remains fixed throughout all evaluations. To estimate the prior score, we adopt the network architecture proposed in~\cite{habi2025learned} and train it over 50 epochs using the AdaW optimizer with a learning rate of 0.0002, 60,000 training samples, and a batch size of 512. Exponential moving averages (EMA) with a decay rate of 0.999 are applied for training stability. The performance is evaluated on 6,000 samples.
We see that the learned BCRB  closely aligns with the analytical BCRB, confirming that the estimated prior score accurately captures the underlying structure of the prior and enables reliable Bayesian FIM computation. In contrast, the analytical CRB, which excludes prior information, yields significantly looser bounds at low transmit power. These results suggest that the scoring ISAC framework offers 
a data-driven way to accurately characterize the localization error bound.

\begin{figure}
    \centering
\includegraphics[width=0.85\linewidth]{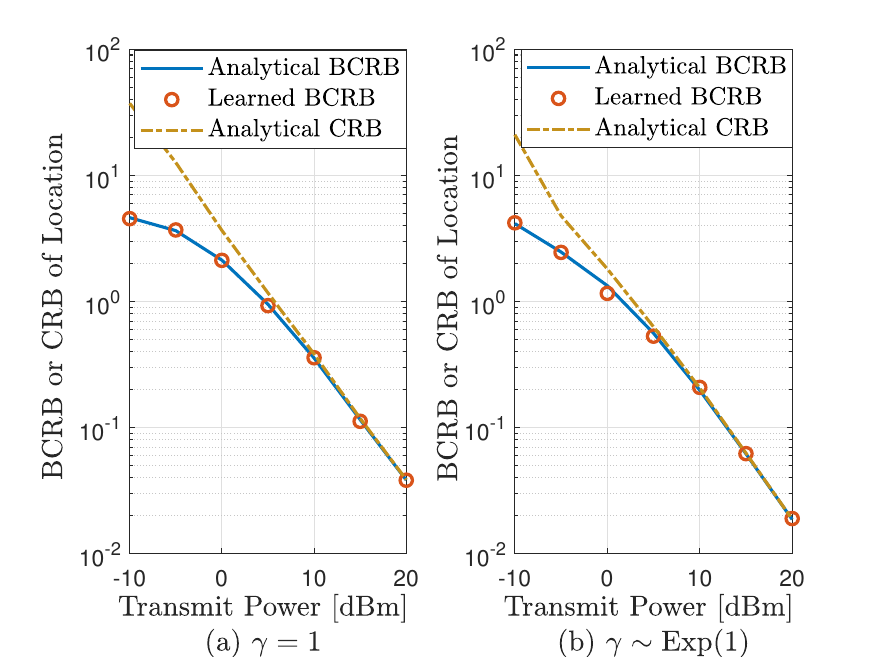}
    \caption{Target localization performance.}
    \label{fig:BCRB_vs_CRB}
\end{figure}

\subsubsection{Unknown Reflection}\label{subsec:BCRBrand}

Consider a general case $\gamma\sim{\rm Exp}(\gamma_0)$ with $\gamma_0=1$. Then, the unknown parameter is $\bm \theta\triangleq[\mathbf{r}^\mathsf{T},\gamma]^\mathsf{T}\in \mathbb{R}^3$ and we rewrite the measurement operator in \eqref{eq:fk} as ${\bm f}(\bm \theta)$. 
With independent $\mathbf{r}$ and $\gamma$, the joint prior is separable as $p(\bm \theta) = p(\mathbf{r})p(\gamma)$, where $p(\mathbf{r}) = \mathcal{N}(\bm \mu_{\rm r}, \sigma_{\rm r}^2 \mathbf{I})$ and $p(\gamma) = {\rm Exp}(\gamma_0)$. 
Therefore, the analytical score of the joint prior $p(\bm\theta)$ is given by
\begin{align}
   & \nabla_{\bm \theta} \log p(\bm\theta)=[\left(\nabla_{\mathbf{r}} \log p(\mathbf{r})\right)^\mathsf{T},\nabla_{\gamma} \log p(\gamma)]^\mathsf{T},
\end{align}
where $\nabla_{\gamma} \log p(\gamma)=-\gamma_0$.
Then, the analytical prior FIM of $\bm \theta$ can be expressed as
\begin{align}
   \mathbf{J}_{{\rm p},\bm{\theta}} = \mathbb{E}_{p(\bm \theta)} \left[ \left( \nabla_{\bm \theta} \log p(\bm\theta) \right) \left( \nabla_{\bm \theta} \log p(\bm\theta) \right)^\mathsf{T} \right] = \begin{bmatrix}
 \frac{1}{\sigma_{\rm r}^2} \mathbf{I} & \mathbf{0} \\
 \mathbf{0} & \gamma_0^2
\end{bmatrix}.
\end{align}
By replacing $\nabla_{\mathbf{r}} \bm{f}(\mathbf{r})$ with $\nabla_{\bm \theta} \bm{f}(\bm \theta)$ in \eqref{eq:analyCRB}, we obtain the likelihood FIM $\mathbf{J}(\bm \theta)$.
The corresponding data FIM is
\begin{align}\label{eq:analyCRB-gamma}
    \mathbf{J}_{{\rm d},\bm \theta}= \mathbb{E}_{p(\bm \theta)}[\mathbf{J}(\bm \theta)]
    \overset{(a)}{=}\begin{bmatrix}
        \mathbf{J}_{{\rm d},\mathbf{r}\mathbf{r}} & \mathbf{J}_{{\rm d},\gamma\mathbf{r}} \\
        \mathbf{J}_{{\rm d},\mathbf{r}\gamma}  & \mathbf{J}_{{\rm d},\gamma\gamma}
    \end{bmatrix},
\end{align}
where (a) is the matrix partitioning operation and $\mathbf{J}_{{\rm d},\mathbf{r}\mathbf{r}}\in\mathbb{R}^{2\times2}$ represents a sub-matrix of $\mathbf{J}_{{\rm d},\bm \theta}\in\mathbb{R}^{3\times3}$. 
Combining prior and likelihood information, the analytical Bayesian FIM is \begin{align}\label{eq:analyBCRB-gamma}
    \mathbf{J}_{{\rm b},\bm{\theta}} = 
    \mathbf{J}_{{\rm p},\bm{\theta}} + \mathbf{J}_{{\rm d},\bm{\theta}} 
    \overset{(a)}{=}\begin{bmatrix}
        \mathbf{J}_{{\rm b},\mathbf{r}\mathbf{r}} & \mathbf{J}_{{\rm b},\gamma\mathbf{r}}\\
        \mathbf{J}_{{\rm b},\mathbf{r}\gamma} & \mathbf{J}_{{\rm b},\gamma\gamma}
    \end{bmatrix},
\end{align}
where (a) follows the same operation as in \eqref{eq:analyCRB-gamma}.

We denote the analytical BCRB and CRB of the target location as $\mathcal{E}_{\rm BCRB}$ and $\mathcal{E}_{\rm CRB}$, respectively. These bounds can be extracted from the respective blocks of the Bayesian FIM $\mathbf{J}_{\rm b,\mathbf{r}}$ and the data FIM $\mathbf{J}_{\rm d,\mathbf{r}}$
by 
\begin{align}\label{eq:schur}
    \mathcal{E}_{\rm BCRB}&={\rm Tr}\left((\mathbf{J}_{{\rm b},\mathbf{r}\mathbf{r}} -\mathbf{J}_{{\rm b},\mathbf{r}\gamma} \mathbf{J}_{{\rm b},\gamma\gamma} ^{-1} \mathbf{J}_{{\rm b},\gamma\mathbf{r}})^{-1}\right),\\
    \mathcal{E}_{\rm CRB}&={\rm Tr}\left((\mathbf{J}_{{\rm d},\mathbf{r}\mathbf{r}} -\mathbf{J}_{{\rm d},\mathbf{r}\gamma} \mathbf{J}_{{\rm d},\gamma\gamma} ^{-1} \mathbf{J}_{{\rm d},\gamma\mathbf{r}})^{-1}\right).
\end{align}
As in Sec.\ref{subsec:BCRBdet}, we consider a scenario where the likelihood is known but the prior for $\mathbf{r}$ and $\gamma$ is unavailable. The Bayesian FIM can be learned using the ISM-based measurement-prior approach. Following the steps in \eqref{eq:analyBCRB-gamma} and \eqref{eq:schur}, the learned Bayesian FIM can then be used to estimate the BCRB $\hat{\mathcal{E}}_{\rm BCRB}$.

In Fig.\ref{fig:BCRB_vs_CRB}(b), we compare the analytical BCRB, analytical CRB, and learned BCRB in the case of $\gamma\sim {\rm Exp}(1)$. The simulation setup is similar to that in Fig.\ref{fig:BCRB_vs_CRB}(a). 
We observe from Fig.\ref{fig:BCRB_vs_CRB}(b) that the learned BCRB closely matches the analytical BCRB.  
Interestingly, both the CRB and BCRB in the case of $\gamma\sim{\rm Exp}(1)$ are slightly lower than those in the case of $\gamma=1$. This may seem counterintuitive at first since the additional randomness in $\gamma$ introduces uncertainty. 
However, in the case of $\gamma\sim{\rm Exp}(1)$, $\mathbb{E}[\gamma^2]=2$ implies that, on average, the received SNR is higher than that in the case of $\gamma=1$. This results in more informative observations, thereby improving the localization performance.

The results in Figs.~\ref{fig:KLD_vs_P_D}-\ref{fig:BCRB_vs_CRB} confirm the ability of the scoring ISAC framework to replicate classical model-based analyses. 
This illustrates its capacity to generalize classical model-based analyses within a data-driven framework. 
Moreover, they highlight the potential of this framework to extend to complex scenarios where only samples from unknown distributions are available.

\section{Conclusion and Future Directions}\label{sec:conclusion}

In this paper, we present a data-driven framework, scoring ISAC, that leverages score-based generative modeling to benchmark fundamental performance metrics in ISAC systems. 
By harnessing the ability of score-based generative models to capture complex distributions, this framework is especially well-suited for realistic ISAC scenarios.
Looking ahead, the growing availability of high-quality ISAC datasets, as summarized in~\cite{zhang2025integrated},
offers valuable opportunities to validate and extend this framework in practical settings. 
Beyond performance evaluation, theoretical metrics also serve as explicit objectives for system optimization. This dual role motivates future integration of score-based generative modeling into algorithm design and system optimization.
Such integration has the potential to overcome challenges posed by nonlinear and uncertain environments that are difficult to model analytically, enabling more efficient and resilient ISAC solutions.

\bibliographystyle{IEEEtran}
\bibliography{reference}

\end{document}